%% file: traffic_games_extended.tex
\def\year{2021}\relax
\documentclass[letterpaper]{article} 
\usepackage{aaai21_arxiv}  
\usepackage{times}  
\usepackage{helvet} 
\usepackage{courier}  
\usepackage[hyphens]{url}  
\usepackage{graphicx} 
\usepackage{natbib}  
\usepackage{caption} 
\author{
	Philipp Geiger, Christoph-Nikolas Straehle
}
\affiliations{
	
	Bosch Center for Artificial Intelligence, Renningen, Germany\\
	philipp.w.geiger@de.bosch.com, christoph-nikolas.straehle@de.bosch.com
	
}
\input{incl/philpreamble}

\newcommand{\suppref}[1]{\ref{#1} in the appendix}
\newcommand{\supprefs}[1]{\ref{#1}}

\newcommand{\paperref}[1]{\ref{#1}}
\newcommand{\Cpaperref}[1]{\Cref{#1}}

\newcommand{\Cpaperrefp}[1]{\Cref{#1}}

\title{Learning game-theoretic models of multiagent trajectories using implicit layers\textsuperscript{*}}

\begin{document}

\newcommand{\insertextendedversionstatement}{Keep in mind that proofs are available in Sec.~\supprefs{sec:proofs} and that the present paper is an extended version of \citep{geiger2021learning}. }

\input{traffic_games_paper_part}

\onecolumn

\appendix

\section*{Appendix}

\input{traffic_games_supp_part}

\bibliography{../manuscript/incl/traffic_games,../manuscript/incl/phil_master,../manuscript/incl/causal_traffic,../manuscript/incl/publications}

\end{document}

%% file: incl/philpreamble.tex
\usepackage[ruled,vlined,linesnumbered]{algorithm2e} 

\usepackage{amsmath,amssymb}
\usepackage{amsthm}
\usepackage{color} 
\usepackage{hyphenat}
\usepackage{bm}  

\usepackage{hyperref}  
\usepackage{cleveref}

\usepackage{tabularx}
\usepackage{boldline}
\usepackage{caption} 
\usepackage{wrapfig}

\usepackage{mathtools}

\usepackage{multirow}

\usepackage{fancyvrb}  

\newtheorem{thm}{Theorem}
\newtheorem{lem}{Lemma}
\newtheorem{prop}{Proposition}

\newtheorem{dfn}{Definition}
\newtheorem{asm}{Assumption}
\newtheorem{rem}{Remark}

\newtheorem{scen}{Scenario}

\Crefname{lem}{Lem.}{Lem.}
\Crefname{prop}{Prop.}{Prop.}
\Crefname{thm}{Thm.}{Thm.}
\Crefname{obs}{Obs.}{Obs.}
\Crefname{exa}{Ex.}{Ex.}
\Crefname{dfn}{Def.}{Def.}
\Crefname{asm}{Assumption}{Assumption}
\Crefname{sett}{Setting}{Setting}
\Crefname{scen}{Scenario}{Scenario}
\Crefname{section}{Sec.}{Sec.}
\Crefname{figure}{Fig.}{Fig.}
\Crefname{rem}{Rem.}{Rem.}
\Crefname{algorithm}{Alg.}{Alg.}

\newcommand{\secref}[1]{\Cref{#1}}

\newcommand{\propref}[1]{\Cref{#1}}
\newcommand{\asmref}[1]{\Cref{#1}}

\DeclareMathOperator*{\R}{\mathbb{R}}

\usepackage{pgfplots}
\usepackage{tikz}
\usetikzlibrary{arrows,backgrounds,fadings}
\usetikzlibrary{bayesnet}
\usetikzlibrary{arrows,arrows.meta}

\newcommand{\stress}{\textbf}

\newcommand{\defi}{\emph}
\newcommand{\socalled}{\emph} 
\newcommand{\todo}[1]{\textcolor{red}{#1}}
\newcommand{\submission}[1]{}

\newcommand{\cando}[1]{}

\newcommand{\reasonwhy}[1]{}
\newcommand{\thought}[1]{}

\newcommand{\idea}[1]{}
\newcommand{\maybe}[1]{}

\newcommand{\nam}[1]{\text{#1}}
\newcommand{\old}[1]{}

\newcommand{\fname}{\mathit}

\usepackage{enumitem}

\newenvironment{citem}{\begin{itemize}[topsep=4pt, partopsep=0pt, itemsep=2pt, parsep=2pt,
		wide=\parindent  
		]}{\end{itemize}}

\newtheorem{CiteTheorem}{Theorem}
\newtheorem{CiteCorollary}{Corollary}
\newtheorem{CiteLemma}{Lemma}
\newtheorem{CiteProposition}{Proposition}
\newenvironment{CTheorem}[1]{\renewcommand*{\theCiteTheorem}{#1}\begin{CiteTheorem}}{\end{CiteTheorem}}

\newenvironment{CLemma}[1]{\renewcommand*{\theCiteLemma}{#1}\begin{CiteLemma}}{\end{CiteLemma}}
\newenvironment{CProposition}[1]{\renewcommand*{\theCiteProposition}{#1}\begin{CiteProposition}}{\end{CiteProposition}}

\newcommand{\utilf}{u}
\newcommand{\utilityf}{\utilf}

\newcommand{\state}{s}

\newcommand{\positions}{Y_0}
\newcommand{\traj}{y}
\newcommand{\ptraj}{\hat{y}}

\newcommand{\trajs}{Y}
\newcommand{\inittraj}{x}

\newcommand{\inittrajs}{X}
\newcommand{\action}{a}
\newcommand{\Action}{A}

\newcommand{\actions}{A}
\newcommand{\game}{\Gamma}
\newcommand{\potf}{\psi}
\newcommand{\gparam}{\theta}
\newcommand{\Gparam}{\Theta}

\newcommand{\agents}{I}
\newcommand{\nagents}{n}

\newcommand{\actionspart}{\tilde{\actions}}
\newcommand{\jacobi}{J}
\newcommand{\hess}{H}
\newcommand{\expli}{g}

\newcommand{\partsindices}{K}
\newcommand{\refpartsindices}{\tilde{K}}

\newcommand{\ai}[2][]{^{#2 #1}}

\newcommand{\shortcommon}{\nam{com}}
\newcommand{\shortown}{\nam{own}}
\newcommand{\shortothers}{\nam{oth}}
\newcommand{\utilfterm}{\utilf}
\newcommand{\finaltime}{T}

\newcommand{\parametrization}{r}

\newcommand{\posx}{v}
\newcommand{\posy}{w}

\newcommand{\pweight}{\hat{q}}

\newcommand{\adim}{{d_{\actions}}}
\newcommand{\sdim}{{d_{\trajs}}}

\newcommand{\pdim}{d_{\gparams}}

\newcommand{\constr}{v}
\newcommand{\multi}{\lambda}

\newcommand{\lne}{\action^*}

\newcommand{\gparams}{\Theta}

\newcommand{\pw}{.49\linewidth}
\newcommand{\ph}{.14\linewidth}

\newcommand{\nrefined}{\tilde{k}}
\newcommand{\trset}{S}

%% file: traffic_games_paper_part.tex
\maketitle

\begin{abstract}
For prediction of interacting agents' trajectories,  we propose an end-to-end trainable architecture that hybridizes neural nets with game-theoretic reasoning, has interpretable intermediate representations, and transfers to downstream decision making.
It uses a net that reveals preferences from the agents' past joint trajectory, and a differentiable implicit layer that maps these preferences to local Nash equilibria, forming the modes of the predicted future trajectory.
Additionally, it learns an equilibrium refinement concept.
For tractability, we introduce a new class of continuous potential games and an equilibrium-separating partition of the action space.
We provide theoretical results for explicit gradients and soundness. 
In experiments, we evaluate our approach on two real-world data sets, where we predict highway drivers' merging trajectories, and on a simple decision-making transfer task.

\end{abstract}

\section{Introduction}
\label{sec:intro}

Prediction of interacting agents' trajectories has recently been advanced by flexible, tractable, multi-modal data-driven approaches.
But it remains a challenge to use them for safety-critical domains with additional verification and decision-making transfer requirements, like automated driving or mobile robots in interaction with humans.
%
%
Towards addressing this challenge, the following seem sensible \emph{intermediate goals}:
(1)~incorporation of well-understood \emph{principles}, prior knowledge and reasoning of the multiagent domain, allowing to generalize well and to transfer to robust downstream decision making;
(2)~\emph{interpretability} of models' latent variables, allowing for verification beyond just \old{empirically} testing the final output;
(3)~theoretical \emph{analysis} of soundness.

In this paper, we take a step towards addressing multiagent trajectory prediction including these intermediate goals, 	while trying to keep as much as possible of the practical strength of data-driven approaches.
For this, we \emph{hybridize neural learning with game-theoretic reasoning} 
-- because
game theory provides well-established explanations 
of agents' behavior based on 
the principle of instrumental rationality, i.e., viewing agents as utility maximizers.
Roughly speaking, we \emph{``fit a game to the observed trajectory data''}.
\maybe{
A natural approach for this is to hybridize neural nets and game theory:
some hidden layers have clear game-theoretic interpretations, e.g. to encode agent's utility functions,
and certain relationships can be constraint/regularized based on game-theoretic principles, in particular observed behavior explained by the rationality (i.e., utility maximization) of agents, as well as prior knowledge, e.g. about agents' utility functions \todo{norms / trade-offs}.
And games naturally come with prescriptive meanings that allow good robust downstream interventional decisions (e.g., by a self-driving car), by taking into account other agent's strategic responses \citep{}.
\todo{verifiability, guarantees for decision making robust to strategic responses.}
}

Along this hybrid direction 
%
one major obstacle -- and a general reason why game theory often remains in abstract settings -- lies in classic game-theoretic solution concepts like the Nash equilibrium (NE) notoriously suffering from computational \emph{intractability}.
As one way to overcome this, we build on \emph{local} NE  \citep{ratliff2013characterization,ratliff2016characterization}.
We combine this with a specific class of games -- \emph{(continuous) potential games} \citep{monderer1996potential} -- for which local NE usually coincide with local optima of a single objective function\maybe{, the potential function}, simplifying search.
Another challenge lies in combining game theory with neural nets in a way that makes the overall model still efficiently \emph{trainable} by gradient-based methods.
%
To address this, we build on \emph{implicit layers} \citep{amos2017optnet,bai2019deep}.
Implicit layers specify the functional relation between a layer's input and its output not in closed form, but only implicitly, usually via an equation.
%
%
%
%
Nonetheless they allow to get exact gradients by ``differentiating through'' the equation, based on the \emph{implicit function theorem}. 
\newcommand{\flexiscript}[2]{{\prescript{#2}{}{#1}}}

\newcommand{\nt}[1]{{\small #1}}
\definecolor{darkred}{rgb}{0.3, 0.1, 0.1}
\definecolor{darkblue}{rgb}{0.1, 0.1, 0.2}
\definecolor{lightblue}{HTML}{EEF2F6}
\begin{figure*}[t]
	\usetikzlibrary{positioning}
	\usetikzlibrary{calc}
	\tikzstyle{stdthick}=[] 
	\tikzstyle{var}=[text width=1.6cm,align=center,color=darkred]
	\tikzstyle{mech}=[rectangle,fill=none,draw=lightgray,align=center,minimum width=1.6cm,minimum height=0.7cm,text width=1.6cm,color=darkblue,stdthick]
	\tikzstyle{stdarrow}=[-{Latex},stdthick]
	\tikzstyle{stdline}=[-{Latex},stdthick] 
	\centering 
	\begin{tikzpicture}[yscale=2,xscale=2.5]
	\node[var] at (0, 0) (pt) {\small past joint trajectory \\ \large $\bm{\inittraj}$ };
	\node[var] at (2, 0) (gp) {\small game parameters \large $\bm{\gparam}$ \maybe{rename game params?}};
	\node[var] at (4, 0) (ln) {\small refined local NEs \large $\bm{(\flexiscript{\lne}{k})_{k \in \refpartsindices}}$}; 
	\node[var,text width=2cm] at (6, 0) (ft) {\small predicted joint trajectories' modes {\large $\bm{\flexiscript{\ptraj}{k}}$} and probabilites {\large $\bm{\flexiscript{\pweight}{k}, k \in \refpartsindices}$}}; 
	\node[var] at (2, -.8) (rp) {\small refined subspace index set \\ \large $\bm{\refpartsindices}$}; 
	
	\node[mech,text width=1.6cm]  at (3, 0) (gs) {\small game solver implicit layer \large $\bm{g}$}; 
	\node[mech]  at (1, 0) (pe) {\small preference revelation net};
	\node[mech]  at (1, -.8) (er) {\small equilibrium refinement net};
	\node[mech]  at (5, 0) (pr) {\small trajectory parametrization \large $\bm{\parametrization}$};
	\node[mech]  at (5, -.8) (es) {\small equilibrium weighting net};
	
	\draw[stdline] (pt) to (pe);
	\draw[stdline,dashed] (pt) to (0, -.8) to (er);
	\draw[stdarrow] (er) to (rp);
	\draw[stdline] (rp) to (3, -.8) to (gs);
	\draw[stdarrow] (pe) to (gp);
	\draw[stdline] (gp) to (gs);
	\draw[stdarrow] (gs) to (ln);
	\draw[stdline] (ln) to (pr);
	\draw[stdarrow] (pr) to (ft);
	
	\draw[stdarrow] (es) to (6, -.8) to (ft);	
	
	\draw[stdarrow] (pr) to (ft);
	
	\draw[path fading=west,dashed] ($ (3, -0.8) + (0.1, 0) $) to ($ (es) - (.6, 0) $);
	\draw[stdline] ($ (es) - (.6, 0) $) to (es);  
	\draw[stdthick,dashed] (ln) to ($ (es) - (1, 0) $);
	
	\node[anchor=west,draw=gray,inner sep=2pt,stdthick] at (-.4, 1.2) (ptp) {\includegraphics[height=2cm]{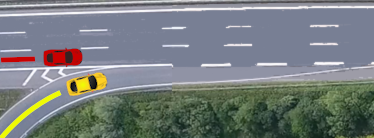}};
	\node[anchor=east,draw=gray,inner sep=2pt,stdthick] at (6.4, 1.2) (ftp) {\includegraphics[height=2cm]{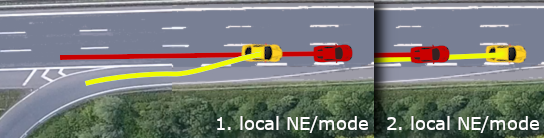}};	
	
	\draw[color=gray,line width=2pt,dotted] (pt) to ($ (ptp.south west) + (.4, 0) $);
	\draw[color=gray,line width=2pt,dotted] (ft) to ($ (ftp.south east) - (.4, 0) $);

	\end{tikzpicture}
	\caption{\emph{Bottom:} Our full architecture (\Cref{sec:archi}). \emph{Top:} Example (stylized): highway merging scenario, where reliable models of (human) driver interaction are key for safe automated driving.  \emph{Top left:} input \emph{$\inittraj$: initial trajectories} of drivers. \emph{Top right:}	
	prediction of \emph{future trajectory $y$}: 
	depicted are \emph{two modes} $\flexiscript{\ptraj}{1}, \flexiscript{\ptraj}{2}$ corresponding to \emph{two local Nash eq.} $\flexiscript{\lne}{1}, \flexiscript{\lne}{2}$: red going first vs.\ yellow first.
	\old{future trajectories $\traj$: depicted are \emph{two modes} corresponding to \emph{two local Nash equilibria} -- yellow merging either after or before red.}
		}
	\label{fig:archi}
\end{figure*}



\paragraph{Main contributions and outline.}  
We propose a modular architecture that outputs a multi-modal prediction of interacting agents' joint trajectory  (where modes are interpretable as local Nash equilibria), from their past trajectory as input (\Cref{sec:approach}). 
The architecture is depicted in \Cref{fig:archi}, alongside the motivating example of highway drivers' merging trajectories.
It builds on the following components:
%
%
\begin{citem}
\item 
a tractable, differentiable \emph{game solver implicit layer} (\Cref{sec:implicitlayerdescription}) with explicit gradient formula, mapping game parameters to local Nash equilibria (\Cref{thm:implicitlayer}).
It is based on a \emph{new class of continuous-action trajectory games} that allow to encode prior knowledge on agents' preferences (\Cref{dfn:commoncoupled}). We prove that they are potential games (\Cref{lem:ispotentialgame}). 
And it builds on an \emph{equilibrium-separating concave partition of the action space} that we introduce to ensure tractability (\Cref{dfn:separating}).
\item Furthermore, the architecture
contains a \emph{neural net that reveals the agents' preferences} from their past, and a \emph{net that learns an equilibrium refinement concept} (\Cref{sec:archi}).
%
%
\end{citem}

This architecture forms a model class where certain latent representations have clear game-theoretic interpretations and certain layers encode game-theoretic principles that help induction (also towards strategically-robust decision-making). At the same time, it has neural net-based capacity for learning, and is end-to-end trainable with analytic gradient formula. Furthermore:

\begin{citem}

\item In \Cref{sec:examplescenarios}, we give two concrete example scenarios that provably satisfy our approach's conditions (\Cref{prop:driving}, etc.).
\item In the experiments reported in \Cref{sec:experiments}, we apply our architecture to prediction of real-world highway on-ramp merging driver interaction trajectories, on one established and one new data set we publish alongside this paper. We also apply it to a simple decision-making transfer task.

\end{citem}

\insertextendedversionstatement 
In what follows, we first discuss related work and introduce setting and background (\Cref{sec:setting}).

\paragraph{Closest related work.}
\label{sec:relatedwork}


\maybe{We are not aware of }
\maybe{while classically, game theory predicts/prescribes agents' actions given knowledge of their preferences and rationality, performing the inverse task (in one way or another) which we do to some extent, is a small but growing area.}
%
Regarding general \emph{multiagent model learning} from observational behavioral data \emph{with game-theoretic components}:
closest related is work by \citet{ling2018game,ling2019large}, who use game solvers as differentiable implicit layers, learning these layers' input (i.e., agents' preferences) from covariates. 
They focus on discrete actions while we address continuous trajectory prediction.
And they use different solution concepts, and do not consider equilibrium refinement.
%
There is further work more broadly related in this direction \citep{kita1999merging,kita2002game,liu2007game,kang2017game,tian2018adaptive,li2018game,fox2018should,camara2018empirical,ma2017forecasting,sun2018probabilistic}, sometimes also studying driver interaction,
but they have no or little data-driven aspects (in particular no implicit layers) and/or use different
approximations to rationality than our local NE, such as level-k reasoning, 
and often are less general than us, often focusing on discrete actions.
More broadly related is multiagent inverse reinforcement learning (IRL) \citep{wang2018competitive,reddy2012inverse,zhang2019non,etesami2020non}, often discrete-action.

\cando{maybe cite more, for instance Multi-Agent Tensor Fusion for Contextual Trajectory Prediction and other stuff, look again at Human Motion Trajectory Prediction: A Survey. look at phikiwi article excerpt/heavily data-driven approaches to multiagent trajectory prediction}
For \emph{multiagent trajectory prediction}, there generally is a growing number of papers on the \emph{machine learning} side, often building on deep learning principles and allowing multi-modality -- but without game-theoretic components. 
Without any claim to completeness, there is work using long-short term memories (LSTMs) \citep{alahi2016social,deo2018convolutional,salzmann2020trajectron}, generative adversarial networks (GANs) \citep{gupta2018social}, and attention-based encoders \citep{tang2019multiple}. \citet{kuderer2012feature} 
uses a partition (``topological variants'') of the trajectory space related to ours. \old{Their model  allows for several macroscopic ``solutions'' (similar to out multitude of equilibria), called topological variants.} 
There is also 
work related to the principle of ``social force'' 
\citep{helbing1995social,robicquet2016learning,blaiotta2019learning}, 
and related rule-based driver modeling approaches \citep{treiber2000congested,kesting2007general}.

Regarding \emph{additional game-theoretic elements}: W.r.t.\ the class of trajectory potential games we introduce (\Cref{dfn:commoncoupled}), the closest related work we are aware of is \citep{zazo2016dynamic} who consider a related class, 
but they do not allow agents' utilities to have differing additive terms w.r.t. their own actions. 
Worth mentioning is further related work based on games (different ones than ours though), but towards pure \emph{control} (not prediction) tasks \citep{peters2020inference,zhang2018addressing,spica2018real,fisac2019hierarchical}. 
\citet{peters2020inference} use a latent variable for the equilibrium selection, similar to our equilibrium weighting.
\old{Different to us, their approach is more based on graphical models and sampling, and does not contain elements like a differentiable game solver with soundness guarantees, or the partition-based \reasonwhy{partition based because implicitly one could also say they learn an eq. refinement} equilibrium refinement learning.}
For further related work see Sec.~\suppref{sec:morerelated}.

%

\section{General Setting, Goals and Background} 
\label{sec:setting}

\thought{fundamental Q: introduce games etc. explicitly, or rather ad-hoc? depends on the audience i want to target. maybe some deep learners would like it.}






We consider \defi{scenes}, each consisting of:
\begin{citem}
\item 
 a set $\agents := \{1, \ldots, \nagents\} \maybe{i \in \agents := \{0, \ldots, \nagents-1\}}$ of agents.
\item
Each agent $i \in \agents$ at each time $t \in [0, \finaltime]$ has an \defi{individual state} $\traj\ai{i}_t \in \R^\sdim$.
They yield an \defi{individual trajectory} $\traj\ai{i} = (\traj\ai{i}_t)_{t \in [0, \finaltime]}$ 
(think of 0 as the present time point and $T$ as the horizon up to which we want to predict).
\item 
And $\traj := ((\traj\ai{1}_t, \ldots, \traj\ai{\nagents}_t))_{t \in [0, \finaltime]} \in \trajs$ 
denotes the agents'  \defi{joint (future) trajectory}. 
\item
We assume that the \emph{past joint trajectory $\inittraj \in \inittrajs$} of the agents until time point 0 is available as side information.
\end{citem}

\old{We assume there is \defi{side information $\inittraj$} available, which, by default, is the agents' \emph{past joint trajectory} until time point 0.}
Now, besides the other goals mentioned in \Cref{sec:intro}, we formulate the \emph{main (passive) predictive problem} as follows:
\begin{citem}
\item 
\emph{goal:} 
in a new scene, predict the future joint trajectory $\traj$
by a list of pairs $(\ptraj, \pweight)$, corresponding to $\traj$'s modes, where each $\ptraj$ is a point prediction of $\traj$, and $\pweight$ the associated probability (for more details and metrics etc., see \Cref{sec:archi}, \ref{sec:experiments}); 
\item 
\emph{given:} (1) the past trajectory $\inittraj$ of that new scene, as well as (2) a training set consisting of previously sampled scenes, i.e., pairs $(\inittraj', \traj'), (\inittraj'', \traj''), \ldots$ of past and future trajectory (discrete-time subsampled of course). (We assume all scenes sampled from the same underlying distribution.) 
\end{citem}


We assume that agent $i$'s (future) trajectory $\traj\ai{i}$ is parameterized by a finite-dimensional vector $\action\ai{i} \in \actions\ai{i} \subseteq \R^\adim$, which we refer to as $i$'s \defi{action}, with $\actions\ai{i}$ the \defi{action space} of $i$. So, in particular, there is a\maybe{ (differentiable)} \defi{(joint) trajectory parameterization $\parametrization : \actions \to \trajs$}, with  $\actions := \actions\ai{1} \times \ldots \times \actions\ai{\nagents}$ the \defi{joint action space}.\thought{\todo{$\traj_0$ is usually known/fixed.}}
Keep in mind that 
$\action = (\action\ai{1}, \ldots, \action\ai{\nagents})$,
$\action\ai{-i}$ means $(\action\ai{1}, \ldots,\action\ai{i-1}, \action\ai{i+1}, \action\ai{\nagents})$ 
and $(\action\ai{i}, \action\ai{-i})$ reads $\action$.
We use games \cite{shoham2008multiagent,osborne1994course} to model our setting. A game specifies the set of agents (also called ``players''), their possible actions and their utility functions.
The following formal definition is slightly tailored to our setting: utilities are integrals over the trajectories parameterized by the actions. 
\begin{dfn}[Game]
	\label{dfn:game}
A \defi{(trajectory) game} consists of: 
the set $\agents$ of agents, and for each agent $i \in \agents$:
the action space $\actions\ai{i} \subseteq \R^\adim$, 
and a utility function $\utilf\ai{i} \colon \actions \to \R$.
We assume $\utilf\ai{i}$, $i \in \agents$, to be of the form 
\[\utilityf\ai{i}(\action) = \int_0^\finaltime \utilityf\ai{i}_t(\traj_{t}) d \mu(t), \]
where $\action \in \actions$, 
$\traj = r(\action)$; 
$\utilityf\ai{i}_t$, $t \in [0, \finaltime]$, are the \defi{stage-wise utility functions},
and $\mu$ is a measure on $[0, \finaltime]$.%
\footnote{This general integral-based formulation contains discrete-time \Cref{sett:driving} as special case with $\mu$'s mass on discrete time points.}
\old{Additionally, we consider a \defi{trajectory parameterization $\parametrization : \actions \to \trajs$} \maybe{by which joint actions determine trajectories}. (That is, agent $i$ by its action $\action\ai{i} \in \actions\ai{i}$ determines its full individual trajectory $\traj\ai{i}$.)}%
\end{dfn}
This game formalizes the agents' ``decision-making \emph{problem}''.
Game theory also provides the ``Nash equilibrium'' as a concept of how rational agents will/should act to \emph{``solve''} the game. Here we use a ``local'' version -- for tractability:
\old{Game theory also provides concepts of how rational agents will/should act to \emph{``solve''} the game. Here we use a ``local'' version for the sake of tractability:}
%


\begin{dfn}[Local Nash equilibrium (NE) \citep{ratliff2016characterization,ratliff2013characterization}] 
	\label{dfn:lne}
Given a game, a joint action $\action \in \actions$ is a \defi{(pure) local Nash equilibrium (local NE)}\old{\todo{maybe:} In this work we only consider pure, not mixed, Nash equilibria and therefore will not explicitly always call them pure.} if there are open sets $S\ai{i} \subset \actions\ai{i}$ such that $\action\ai{i} \in S\ai{i}$ and for each $i$,
\[ \utilf\ai{i}(\action\ai{i}, \action\ai{-i}) \geq \utilf\ai{i}(\action\ai[\prime]{i}, \action\ai{-i}),  \] 
for any $\action\ai[\prime]{i} \in S\ai{i}$.%
\footnote{I.e., no agent can improve its utility by unilaterally and locally deviating from its action in a local NE -- a ``consistency'' condition.
}
\old{If the above inequality is strict, we call $a$ a \defi{strict (pure) local NE}. \cando{do we need strict? maybe drop it}}
If $S\ai{i} = \actions\ai{i}$ for all $i$, then $a$ is called a \defi{(pure, global) NE}.
\end{dfn}

The following type of game 
can reduce finding local NE to finding local optima of a single objective (``potential function''), allowing for tractable gradient ascent-based search.

\old{, making it tractable, via gradient descent and related methods.\footnote{The idea behind ... non-adverserial(?) ... classical example are congestion games.}}

\begin{dfn}[Potential game \citep{monderer1996potential}] \maybe{,rosenthal1973class}
	\label{dfn:potgame}
A game is called an \defi{(exact continuous) potential game}, if there is a so-called \defi{potential function} $\potf$ such that,
for all agents $i$, all actions $\action\ai{i}, \action\ai{i'}$ and remaining actions $\action\ai{-i}$,
\[
\utilf\ai{i}(\action\ai{i'}, \action\ai{-i}) - \utilf\ai{i}(\action\ai{i}, \action\ai{-i}) = \potf(\action\ai{i'}, \action\ai{-i}) - \potf(\action\ai{i}, \action\ai{-i}).
\]
\end{dfn}



Let us also give some neural net-related background:

\begin{rem}[Implicit layers \citep{amos2017optnet,amos2018differentiable,bai2019deep,el2019implicit}]
	\label{rem:il}
	Classically, one specifies a neural net layer by specifying the functional relation between its input $v$ and output $w$ \emph{explicitly}, in \emph{closed form}, $w = f(v)$, for some function $f$ (e.g., a softmax).
	The idea of \defi{implicit layers} is to specify the relation \emph{implicitly}, usually via an equation $h(v, w) = 0$ (coming from, e.g., a stationarity condition of an optimization or dynamics modeling problem). \old{(While building on long-standing ideas and previous developments in this direction, e.g. for optimization and differential equation problems \citep{chen2018neural,amos2017optnet}, the general idea \citet{bai2019deep,el2019implicit} (who in particular propose to replace deep sequential nets that converge against a fixed point / equilibrium by implicit layers ... ``well-posedness property'') is still rather new.)}
	To ensure that this specification is indeed useful in prediction and training, there are two important requirements:
	(1) the equation has to determine a unique, tractable function $f$ that maps $v$ to $w$, and
	(2) $f$ has to be differentiable, ideally with explicitly given analytic gradients.
	\cando{vlt can man diesen abschnitt auch in sec3.1 moven, da das ja eh nicht formal ist und kurz hat man es ja schon in der intro eingefuehrt}
	\end{rem}

\section{General Approach With Analysis} 
\label{sec:approach}

We now describe our general approach.
It consists of (a)~a game-theoretic model and differentiable reasoning 
about how the agents behave (\Cref{sec:ourgame}), and (b)~a neural net architecture that incorporates this game-theoretic model/reasoning as an implicit layer and combines it with learnable modules, with tractable training and decision-making transfer abilities (\Cref{sec:archi}).



\subsection{Common-Coupled Games, Equilibrium- Separation and Induced Implicit Layer} 
\label{sec:implicitlayerdescription}
\label{sec:ourgame}


For the rest of the paper, let $(\game_\gparam)_{\gparam \in \gparams}$, $\gparams \subseteq \R^{\pdim}$, be a \emph{parametric family of trajectory games} (\Cref{dfn:game}).
%
First let us introduce the following type of trajectory game to strike a balance between adequate modeling and tractability:
\begin{dfn}[Common-coupled game]
\label{dfn:commoncoupled}
\label{asm:normcoupled}
We call $\game_\gparam$ a 
\defi{common-coupled(-term trajectory) game}%
, if the stage-wise utility functions (\Cref{dfn:game}) have the following form, for all agents $i \in \agents$, $t \in [0, \finaltime]$:
%
\begin{align}
\label{eqn:defcoupledutil}
\utilityf\ai{i, \gparam}_t(\traj_{t}) &= \utilfterm\ai{\shortcommon, \gparam}_t(\traj_{t}) + \utilfterm\ai{\shortown, i, \gparam}_t(\traj\ai{i}_{t}) + \utilfterm\ai{\shortothers, i, \gparam}_t(\traj\ai{-i}_{t}), 
\end{align}
\old{where $\state_t$ denotes the state at time $t$, where we assume that the state can be separated and for each individual agent just depends on its own past actions (as naturally the case in certain physical spaces) and the state should be thought of as the state based on the trajectory (up to $t$) in the sense of physics, but we will have several slightly different specific versions of it in the paper, and }
\maybe{or maybe just replace $x$'s by state $s$'s and say that states are separable functions of past}
where
$\traj = \parametrization(\action)$ (action parameterizes trajectory, \Cref{sec:setting}),
$\utilfterm\ai{\shortcommon, \gparam}_t$ is a term that depends on all agents' trajectories and is common between agents, 
$\utilfterm\ai{\shortown, i, \gparam}_t$ and $\utilfterm\ai{\shortothers, i, \gparam}_t$ are terms that only depend on agent $i$'s trajectory, or all other agents' trajectories, respectively, and may differ between agents.
%
\old{We call such a game a \defi{convention-coupled -- same-coupled -- norm-coupled -- common-coupled(-additive-term)  (trajectory) game}.}
\end{dfn}


%

%
%


Common-coupled games adequately approximate many multiagent trajectory settings where agents trade off (a) social norms and/or common interests (say, traffic rules or the common interest to avoid crashes), captured by the common utility term $\utilfterm\ai{\shortcommon, \gparam}_t$, against (b) individual inclinations related to their own state, captured by the terms $\utilfterm\ai{\shortown, i, \gparam}_t$.
It is non-cooperative, i.e., utilities differ, but more on the cooperative than adversarial end of games. 
\old{Reviewer 5: The common-coupled game assumes that the only component of my utility that depends jointly on my position and the other cars' positions is shared among all drivers. This is briefly justified by gesturing toward crashes as being bad for everyone. I would suggest a little more justification is needed here.}
For tractability we can state:

\begin{lem}
\label{lem:ispotentialgame}
If $\game_\gparam$ is a common-coupled game, then it is a potential game with the following potential function, where, as usual, $\traj = \parametrization(\action)$:
\[
\potf(\gparam, \action) = \int_0^\finaltime \utilfterm\ai{\shortcommon, \gparam}_t(\traj_{t}) + \sum_{i \in \agents} \utilfterm\ai{\shortown, i, \gparam}_t(\traj^i_{t}) d \mu(t) .
\] 

\end{lem}


Note that this implies existence of NE, given continuity of the utilities and compactness \citep{monderer1996potential}.%
We now show how the mappings from parameters $\gparam$ to local NE of the game $\game_\gparam$ can be soundly defined, tractable and differentiable, so that we can use this game-theoretic reasoning\footnote{Here, we mean reasoning in the sense of drawing the ``local NE conclusions'' from the game $\game_\gparam$, due to the principle of rationality. More generally, game-theoretic reasoning also comprises equilibrium refinement/selection (\Cref{sec:archi}).} 
as one implicit layer (\Cref{rem:il}) in our architecture.
For this, a helpful step is to (tractably) partition the action space into subspaces with exactly one equilibrium each -- if the game permits this.
For the rest of the paper, let $(\actionspart_k)_{k \in \partsindices}$ be a finite \emph{collection of subspaces of $\actions$}, i.e., $\actionspart_k \subseteq \actions$.

\begin{dfn}[Equilibrium-separating action subspaces]
\label{dfn:separating}
For a common-coupled game $\game_\gparam$, we call the action subspace collection $(\actionspart_k)_{k \in \partsindices}$ \defi{equilibrium-separating (partition)} if, for all $k \in \partsindices$ and $\gparam \in \gparams$, the game's potential function $\potf (\gparam, \cdot)$ is strictly concave on $\actionspart_k$.%
\footnote{%
We loosely speak of a \emph{partition} of $\actions$, but we do not require to cover the full $\actions$, and we allow overlaps, so it is not a partition in the rigorous set-theoretic sense\old{, and not necessarily cover the whole space}. NB: The subspaces also have the interpretation as \emph{macroscopic/high-level joint action} of the agents: for instance, which car goes first in the merging scenario in \Cref{fig:archi}.}
\old{there is exactly one local NE of $\game_\gparam$ in each $\actionspart_k$.}
\thought{problem: this is more an assumption about potf instead of collection ......}
\end{dfn}
As a simplified example, a first partition towards equilibrium-separation in the highway merging scenario of \Cref{fig:archi} would be into two subspaces: (1)~those that result in \emph{joint trajectories where the red car goes first} and (2)~those where \emph{yellow goes first}. 
More details follow in \Cref{sett:driving}. 

\reasonwhy{actually all we need for the thm is parts with exactly one EQ (and compactness or so?). do not need anything like non-overlapping. but then partition is not the right word. or maybe stick with partition but say that it can in fact be a pratition of only a subspace becaues we need some gaps between the parts? maybe wait until i have written down the proofs for the subsettings.}

%
Keep in mind that the equation $\nabla_{\action} \potf (\gparam, \action) {=} 0$ is a necessary condition for $\action$ to be a local NE of $\game_\gparam$ (for interior points), since local optima of the potential function correspond to local NE.
This equation induces an implicit layer:

\cando{i think the NE/optima is now already stated three times, so maybe can drop it somewhere}


\begin{asm}
\label{asm:diffetc}
Let $\game_\gparam$ be a common-coupled game. 
Let $(\actionspart_k)_{k \in \partsindices}$ be equilibrium-separating subspaces for it,
and let all $\actionspart_k, k \in \partsindices$ be compact, \old{and convex } given by the intersection of linear inequality constraints.
On each subspace $\gparams \times \actionspart_k, k \in \partsindices$, 
let $\game_\gparam$'s potential function $\potf$ be continuous.
\end{asm}


\begin{thm}[Games-induced differentiable implicit layer] 
	\maybe{maybe have to thm parts: one fore the case taht we can guarantee unique optimum (essentialy this might be continuity claim plus first point) and one only regaring the gradient, given diff (the second bullet)}
\label{prop:gradient}
\label{thm:implicitlayer}
Let \Cref{asm:diffetc} hold true.%
\footnote{Note: (1) (Parts of) this theorem translate to general potential games, not just common-coupled games. 
(2) For tractability/analysis reasons we consider the simple deterministic game form of \Cref{dfn:game} instead of, say, a Markov game -- which we leave to future work.
(3) Our framework may still be applicable if assumptions like concavity, which is quite strong, are relaxed. However, deriving guarantees may become arbitrarily hard.
}
Then, for each $k \in \partsindices$, there is a continuous mapping $\expli_k : \gparams \to \actionspart_k$, such that for any $\gparam \in \gparams$, if $\expli_k(\gparam)$ lies in the interior of $\actionspart_k$, then 

\begin{citem}
\item $\expli_k(\gparam)$ is a local NE of $\game_\gparam$,
\item $\expli_k(\gparam)$ is given by the unique argmax\footnote{Due to the concavity assumption, we can use established tractable, guaranteeably sound algorithms to calculate this argmax.} 
of $\potf (\gparam, \cdot)$ on $\actionspart_k$, with $\potf$ the game's potential function (\Cref{lem:ispotentialgame}), 
\item  
$\expli_k$ is continuously differentiable in $\gparam$ with gradient 
\[\jacobi_\gparam \expli_k (\gparam) =  - \left( \hess_\action \potf (\gparam, \action) \right)^{-1} \jacobi_{\gparam} \nabla_{\action} \potf (\gparam, \action), \] 
whenever $\potf$ is twice continuously differentiable on an open set containing $(\gparam, \action)$, for $\action = \expli_k (\gparam)$,
\maybe{\item if $\expli(\gparam)$ lies on (exactly) one constraining hyperplane of $\actionspart_k$ defined by orthogonal vector ... and with strictly positive Lagrange multiplier  ...., then ...}
where $\nabla, \jacobi$ and $\hess$ denote gradient, Jacobian and Hessian, respectively.
\end{citem}
\maybe{
In particular, for any $k_1, \ldots, k_\ell$, ............
\todo{Furthermore, if additionally assumption A2 holds, then ...}
%
%
\begin{citem}
	\item if A1 holds, then, if $\expli(\gparam)$ lies in the interior, then the gradient at $\gparam$ is given by
	\item if A2 holds and if $\expli(\gparam)$ lies in the interior, or if (exactly) the $m$th constraint is active and $\lambda* > 0$, then the gradient
\end{citem}
%
%
%
}
\end{thm}

The specifics of how the $g_k$ of \Cref{thm:implicitlayer} form an implicit layer
will be discussed in \Cref{sec:archi}. 

\paragraph{Remark on boundaries.}
There remain several questions: e.g., whether the action space partition introduces ``artificial'' local NE at the boundaries of the subspaces; and also regarding what happens to the gradient if $\expli_k(\gparam)$ lies at the boundary of $\actions$ or $\actionspart_k$.
Here we state a preliminary answer\footnote{Note that similar results have already been established, in the sense of constrained optimizers as implicit layers \citep{amos2017optnet}, but we give the precise preconditions for our setting. See also Sec.~\suppref{sec:morerelated}.
Moreover, note that under the conditions of this lemma, i.e., when $\expli_k (\gparam)$ lies at the boundary, then the above gradient $\jacobi_{\gparam} \expli_k (\gparam)$ in fact often becomes zero, which can be a problem for parameter fitting.
} to the latter:
\begin{lem}
\label{lem:gradbound}
Assume \asmref{asm:diffetc} and that
$\potf$ is twice continuously differentiable on a neighborhood of $\gparams {\times} \actionspart_k, k {\in} \partsindices$. 
If $a{=}\expli_k(\gparam)$ lies on exactly one constraining affine hyperplane of $\actionspart_k$, defined by orthogonal vector $\constr$, with multiplier $\lambda$ and optimum $\lambda^* {>} 0$ of $\potf(\gparam, \action)$'s Lagrangian (details see proof), then
$\jacobi_{\gparam} \expli_k (\gparam)$
is the upper left $\nagents \adim {\times} \nagents \adim$-submatrix of

$-\left(\begin{array}{cc}\hess_a \potf(\gparam, \action)  & \constr \\ \lambda^* \constr^T & 0 \\ \end{array} \right)^{-1}  \left( \begin{array}{c} \jacobi_{\gparam} \nabla_\action \potf(\gparam, \action) \\ 0 \end{array} \right) 
$.
\end{lem}

\paragraph{Remark on identifiability.}
Another natural question is whether the game's parameters are \emph{identifiable} from observations,
and, especially, whether the $\expli_k$ are invertible. \maybe{or $\expli_k$? i guess if one $\expli_k$ is invertible this means the whole is invertible ... what do we want in the end? i guess id from the final output ... need to think about this ... what if subspaces overlap?}
While 
difficult to answer in general, we investigate this for one scenario in Sec.~\suppref{sec:pedestriandetails}.

\subsection{Full Architecture With Further Modules, Tractable Training and Decision Making}
\label{sec:archi}
\label{sec:decisionmaking}

Now for the overall problem of mapping past joint trajectories $\inittraj$ to predictions of their future continuations $\traj$, 
we propose the architecture depicted in \Cref{fig:archi} alongside a training procedure.
\old{to address the general \todo{multiagent trajectory modeling} problem (\Cref{sec:problem})}
We call it \defi{trajectory game learner (TGL)}. 
(Its forward pass is explicitly sketched in Alg.~\supprefs{alg:tglf} in Sec.~\suppref{sec:tgldetails}.)
It contains the following modules (here we leave some of the modules fairly abstract because details depend on size of the data set etc.; for one concrete instances see the experimental setup in \Cref{sec:experiments}), which are well-defined under \Cref{asm:diffetc}:

\begin{citem}
\item \stress{\defi{Preference revelation net}:}
It maps the past 
joint trajectory $\inittraj \in \inittrajs$ to the inferred game parameters $\gparam \in \gparams$ (encoding agents preferences).\footnote{In a sense, this net is the inverse of the game solver implicit layer on $\inittraj$, but can be more flexible.}
For example, this can be an LSTM.



\item \stress{\defi{Equilibrium refinement net}:}
This net maps the past joint trajectory $\inittraj \in \inittrajs$ to a subset $\refpartsindices \subset \partsindices$(we encode $\refpartsindices$ e.g.\ via a multi-hot encoding), with $| \refpartsindices | = \nrefined $, 
for $\nrefined$ arbitrary but fixed.
This subset $\refpartsindices$ selects a \emph{subcollection} $(\actionspart_k)_{k \in \refpartsindices}$ of the full equilibrium-separating action space partition $(\actionspart_k)_{k \in \partsindices}$ (introduced in \Cref{sec:implicitlayerdescription}, \Cref{dfn:separating}).
This directly determines a \emph{subcollection of local NE of the game $\game_\gparam$}, denoted by $(\flexiscript{\lne}{k})_{k \in \refpartsindices}$ -- those local NE that lie in one of the subspaces $\actionspart_k, k \in \refpartsindices$.%
\footnote{To be exact, in rare cases it can happen that some of these local NE are ``artificial'' as discussed in \Cref{sec:implicitlayerdescription}.} 
The purpose is to narrow down the set of all local NE to a ``refined'' set of local NE that form the ``most likely'' candidates to be selected by the agents. 
\footnote{In game theory, ``equilibrium refinement concepts'' mean hand-crafted concepts that narrow down the set of equilibria of a game (for various reasons, such as achieving ``stable'' solutions) \cite{osborne1994course}. \maybe{shoham2008multiagent,}
For us, the ``locality relaxation'' makes the problem of ``too many'' equilibria particularly severe, since the number of \emph{local} NE can be even bigger than global NE; it can grow exponentially in the number of agents in our scenarios.
}
The reason why we not directly output the refined local NE (instead of the subspaces) is to simplify training (details follow).
%
%
As a simple example, take a feed forward net with softmax as final layer to get a probability distribution over $\partsindices$, and then take the $\nrefined$ most probable $k \in \partsindices$ to obtain the set $\refpartsindices$.

\item \stress{\defi{Game solver implicit layer}%
\footnote{NB: Here, the implicit layer does not have parameters. Generally, implicit layers with parameters can be handled similarly.}
$\expli := (\expli_k)_{k \in \refpartsindices}$:}
It maps the revealed game parameters $\gparam \in \gparams$ together with the refined $\refpartsindices$ to the refined subcollection $(\flexiscript{\lne}{k})_{k \in \refpartsindices}$ of local NE%
\footnote{At first sight, local NE are a poorer approximation to rationality than global NE, and are mainly motivated by tractability. However, we found that in various scenarios, like the highway merging, local NE do seem to correspond to something meaningful, like the intuitive modes of the distribution of joint trajectories. \maybe{, e.g., which car goes first in \Cref{fig:archi}.} NB: Generally, we do not consider humans as fully (instrumentally) rational, but we see (instrumental) rationality as a useful approximation.}
(described in the equilibrium refinement net above).
This is done by performing, for each $k \in \refpartsindices$, the concave optimization over the subspace $\actionspart_k$:
\[
\flexiscript{\lne}{k} = \expli_k(\gparam) = \arg\max_{\action \in \actionspart_k} \potf (\gparam, \action) ,
\]
based on \Cref{thm:implicitlayer}.
See also Line~\supprefs{line:f} to \supprefs{line:fe} in Alg.~\supprefs{alg:tglf} in Sec.~\suppref{sec:tgldetails}.


\item \stress{\defi{Equilibrium weighting net}:}
It outputs probabilities $(\flexiscript{\pweight}{k})_{k \in \refpartsindices}$ over the refined equilibria, and thus \emph{probabilities of the modes of our prediction} (introduced in \Cref{sec:setting}).
We think of them as the probabilities of the mixture components in a mixture model,
but leave the precise metrics open.
%
%
As input, in principle the variables $\gparam, (\flexiscript{\lne}{k})_{k \in \refpartsindices}$ are allowed, plus possibly the agents' utilities attained in the respective equilibrium. And one can think of various function classes, for instance a feed forward net with softmax final layer.
Its purpose is to (probabilistically) learn agents' ``equilibrium selection'' mechanism considered in game theory.\footnote{``Equilibrium selection'' \citep{harsanyi1988general} refers to the problem of which \emph{single} equilibrium agents will end up choosing if there are multiple -- possibly even after a refinement.}

\item \stress{Trajectory parameterization $\parametrization$:} This is the pre-determined parameterization from \Cref{sec:setting}: it maps each local NE's joint action $\flexiscript{\lne}{k}$ to the corresponding joint trajectory $\flexiscript{\ptraj}{k}$ that results from it,
corresponding to \emph{mode $k$ of the prediction},
where $k \in \refpartsindices$ are the indices of the refined equilibria.
\end{citem}


\maybe{$a$ to $\lne$ everywhere?}


%
%
%
%
%




\paragraph{Training and tractability.}
\label{sec:implementationetc}

Training of the architecture in principle happens as usual by fitting it to past scenes in the training set,
sketched in Alg.~\supprefs{alg:tglt} in Sec.~\suppref{sec:tgldetails}. The implicit layer's gradient for backpropagation is given in \Cref{thm:implicitlayer}.
By default, we take the mean absolute error (MAE) averaged over the prediction horizon $[0,T]$ (see also \Cref{sec:experiments}). 
%
%
%
Note that the full architecture -- all modules plugged together -- is not differentiable, because the equilibrium refinement net's output is discrete. 
However, it is easy to see that (1) the equilibrium refinement net and (2)  the rest of the architecture can be trained \emph{separately} and are both differentiable themselves\old{ ... essentially $p(\traj | \inittraj) = \sum_k p(\traj | k, \inittraj) p(k | \inittraj) = 0 + \ldots + .........$}: in training, for each sample $(\inittraj, \traj)$, we directly know which subspace $\traj$ lies in, 
so we first only train the equilibrium refinement net with this subspace's index $k$ as target,
and then train the full architecture with the equilibrium refinement net's weights fixed.\footnote{Therefore we loosely refer to the full architecture as ``end-to-end \emph{trainable}'', not ``end-to-end \emph{differentiable}''.}
\footnote{On a related note, we learn the common term's parameter $\gparam$ (see \eqref{eqn:defcoupledutil}) as shared between all scenes, 
	while the other parameters are predicted from the individual's sample past trajectory.}

Observe that in training there is an outer (weight fitting) and an inner (game solver, i.e., potential function maximizer, during forward pass) optimization loop, so their speed is crucial. For the game solver, 
we recommend quasi-Newton methods like L-BFGS, because this 
is possible due to the subspace-wise concavity of the potential function (\Cref{asm:diffetc}). For the outer loop, we recommend recent stochastic quasi-Newton methods \citep{wang2017stochastic,li2018implementation}. 

%

\paragraph{Transferability to decision making.}
Once the game $\game_\gparam$'s parameters $\gparam$ are learned (for arbitrary numbers of agents) as described above, it does not just help for \emph{prediction} -- i.e., a model of how an \emph{observed} set of strategic agents \emph{will} behave -- but also for \emph{prescription}. This means (among other things) that it tells how a newly introduced agent \emph{should} decide to \emph{maximize its utility}, while aware of how the other agents respond to it based on their utilities in $\game_\gparam$ (think of a self-driving car entering a scene with other -- human -- drivers).%
\footnote{This is the general double nature of game theory -- predictive and prescriptive \citep{shoham2008multiagent}.}
Note: the knowledge of $\game_\gparam$ cannot resolve the remaining equilibrium selection problem (but the equilibrium weighting net may help).
For an example see \Cref{sec:experimentsdecision}.

\section{Concrete Example Scenarios With Analysis}
\label{sec:examplescenarios}

\begin{figure}[!t]
	\centering
	\includegraphics[width=.45\linewidth]{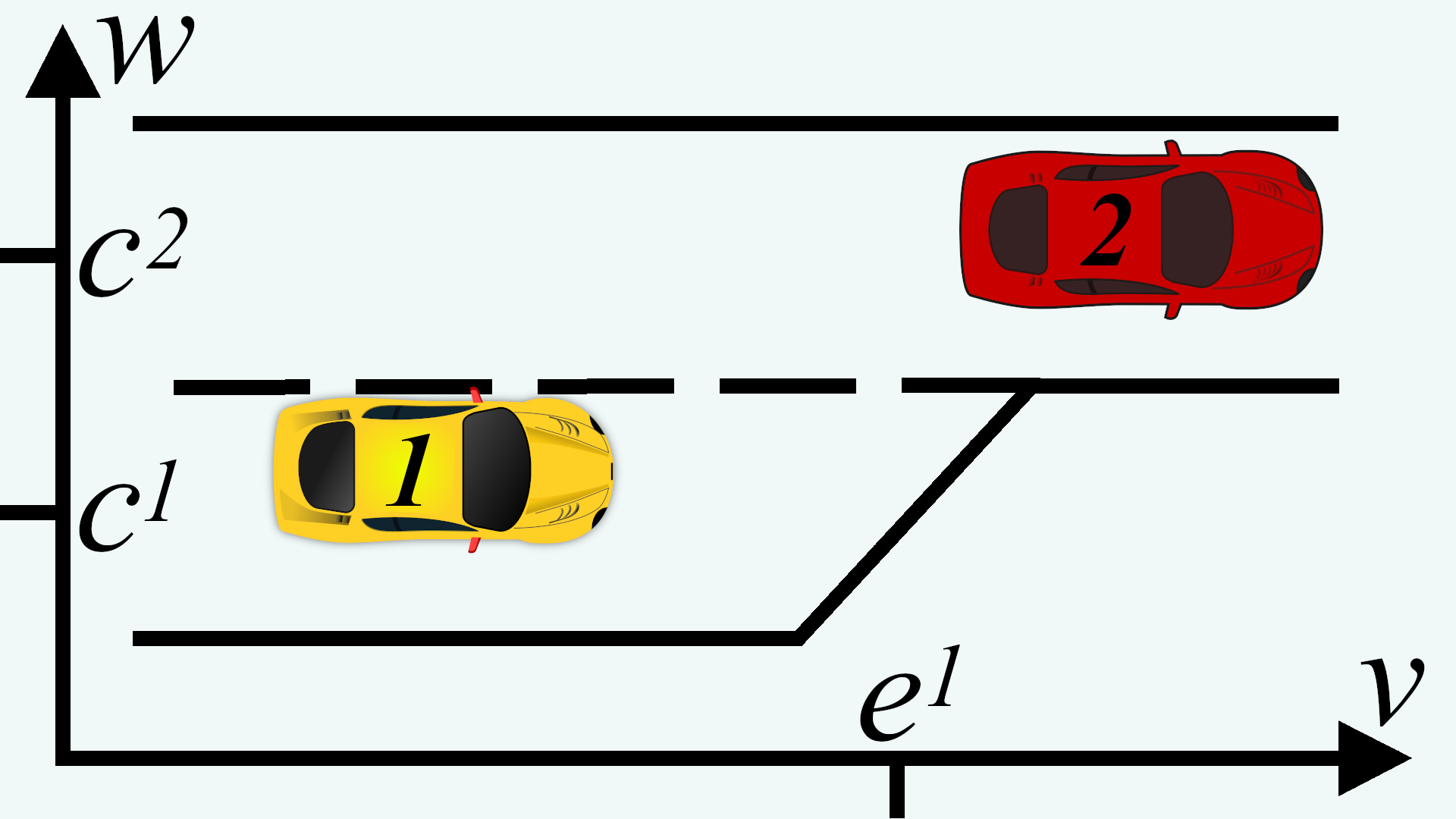}
	\hspace{.4cm}
	\includegraphics[width=.45\linewidth]{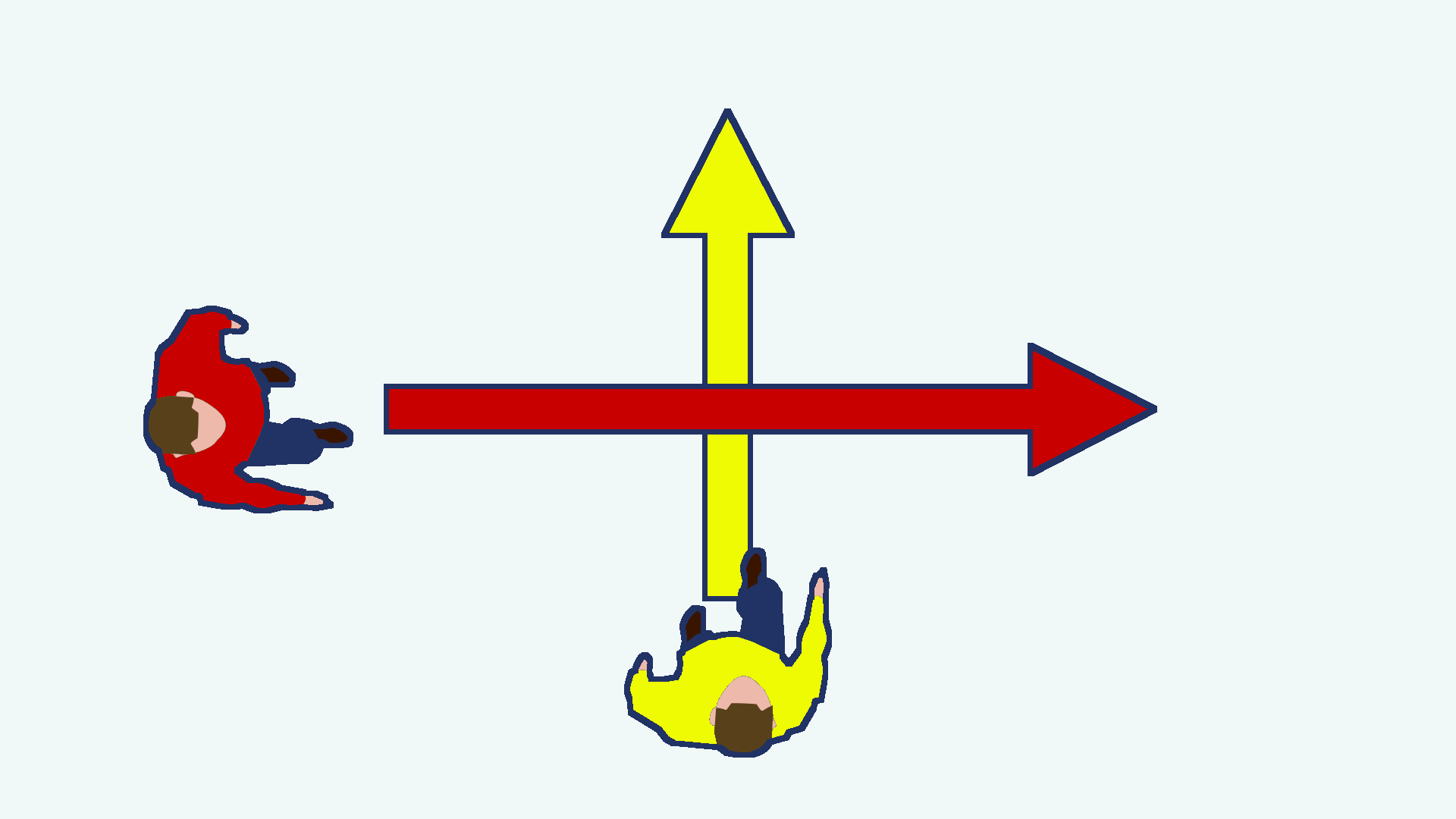}
	\caption{\emph{Left:} Simple illustration of \Cref{sett:driving}'s variables. \emph{Right:} Illustration of simplistic pedestrian encounter scenario (Sec.~\suppref{sec:pedestriandetails}).}
	\label{fig:drivervars}
	\label{fig:pedestrians}
\end{figure}

We give two examples of settings
alongside games and action space partitions
that provably fulfill the conditions for our general approach (\Cref{sec:approach}) to apply.
%
\old{
\todo{While our general framework does not require convexity of the utilities, in the following we make heavy use of convexity beacuse it drastically simplifies analysis.}
\todo{Several points where left open in the general approach: in particular, can we find tractable equilibrium-separating partitions? (Otherwise the equilibrium finding may sort of be just shifted to the partition finding.)}
\todo{Empirically we also found other settings where it works; however, analysis can get increasingly complex and here we focus on proved ones.}
}
%
%
%
%
%
%
First we consider a scenario that captures various non-trivial driver interactions like overtaking or merging at on-ramps.
Essentially, it consists of a straight road section with multiple (same-directional) lanes, where some lanes can end within the section. \Cref{fig:archi} and \ref{fig:drivervars} (left) are examples. 
\thought{what about gegenrichtung? -- in a sense this is fine if there are no mixes betweeen directions allowed then we can simply separate it}
This setting will be used in the experiments (\Cref{sec:experiments}).

\begin{scen}[Multi-lane driver interaction]
\label{sett:driving}
\stress{Setting:}
The set of possible individual states, denote it by $\positions$, is of the form $[b,c] \times [d, e]$ -- positions on a road section.
There are $m$ parallel lanes (some of which may end), parallel to the x-axis.
Agent $i$'s action $\action\ai{i} \in \actions\ai{i}$ is given by the sequence of planar (i.e., 2-D) positions denoted $(\posx\ai{i}_t, \posy\ai{i}_t) \in \positions, t=0, \ldots, \finaltime$, but not allowing backward moves (and possibly other constraints). 
Define the states $\traj\ai{i}_t := ((\posx\ai{i}_t, \posy\ai{i}_t), (\posx\ai{i}_{t-1}, \posy\ai{i}_{t-1}), (\posx\ai{i}_{t-2}, \posy\ai{i}_{t-2}))$.\footnote{We do this state augmentation so that utilities can also depend on velocity/acceleration (not just position) while still rigorously fitting into \Cref{dfn:game}.
When calculating prediction errors for $\ptraj$, only the position component is considered.}
And let $\traj\ai{i}$ be the linear interpolation.
\old{\footnote{\todo{To rigorously fit into our trajectory game definition \Cref{dfn:game}, actually we have to perform state augmentation and define $\parametrization$ accordingly: $\traj\ai{i}_t =((\posx\ai{i}_t, \posy\ai{i}_t), (\posx\ai{i}_{t-1}, \posy\ai{i}_{t-1}), (\posx\ai{i}_{t-1}, \posy\ai{i}_{t-1}))$.
This is because the utility terms $\utilfterm\ai{i, \gparam}_t$ in \Cref{eqn:driver} also depend on velocity/acceleration, not just current position.
But the prediction errors are only calculated w.r.t.\ the position component ...... This augmentation would not be part of the prediction $\ptraj$ though, obviously.
}}}
\stress{Game:} Let, for $t=0, \ldots, T$, the stage utilities of agent $i$ in the game $\game_\gparam$ be the following sum of terms for distance between agents, 
distance to center of lane, 
desired velocities, 
acceleration penalty, 
and 
end of lane overshooting penalty, respectively:%
\footnote{Note that the \emph{invariance over time} of the utility terms, as we assume it here, is a key element of how rationality principles can give \emph{informative priors}.}
\begin{subequations}
\label{eqn:driver}
\begin{align}
&\utilfterm\ai{i, \gparam}_t(\traj^i_{t}) = - \gparam\ai{\nam{dist}} \sum 
\frac{1}{ | \posx\ai{j'}_t - \posx\ai{j}_t | {+} \zeta} - \gparam\ai{\nam{cen}, i}_t (\posy\ai{i}_t {-} c\ai{i}_t     )^2 \\
&- \gparam\ai{\nam{vel}, i}_t (\delta \posx\ai{i}_t - \gparam\ai{\nam{v}, i}     )^2 -  \gparam\ai{\nam{velw}, i} (\delta \posy\ai{i}_t )^2 - \gparam\ai{\nam{acc}, i}  ( \delta^2 \posx\ai{i}_t )^2 \\ &- \gparam\ai{\nam{end}, i} \max(0, \posx_t - e\ai{i}_t),
\end{align}
\end{subequations}
where 
the sum ranges over all $(j, j')$ such that driver $j$ is right before $j'$ on the same lane;\reasonwhy{we do have to range over all cars here and not just the neighbors of i because of the potential game requirement}
$\zeta > 0$ is a constant,
$c\ai{i}_t$ is the respective center of the lane, 
$\delta \posx\ai{i}_t$ means velocity along lane, 
$\delta \posy\ai{i}_t$ means lateral velocity,
$\delta^2 \posx\ai{i}_t$ means acceleration (vector), 
$e\ai{i}_t$ is the end of $i$'s lane, if it ends, otherwise $-\infty$;
furthermore, $\mu$ is the counting measure on $\{0, \ldots, \finaltime \}$ (i.e., discrete). 
and $\gparam {=} (\gparam\ai{\nam{dist}}, \gparam\ai{\nam{cen}, i}_{[0:T]}, \gparam\ai{\nam{vel}, i}_{[0:T]}, \gparam\ai{\nam{v}, i}, \gparam\ai{\nam{velw}, i}, \gparam\ai{\nam{acc}, i}, \gparam\ai{\nam{end}, i})_{i \in \agents}$.\footnote{We allow some of the weights to vary with $t$ to add some flexibility. In the experiments (\Cref{sec:experiments}), we use ``terminal'' costs only; more specifically $\gparam\ai{\nam{vel}, i}_t = 0$ for $0 \leq t \leq T-6$ and $\gparam\ai{\nam{cen}, i}_t = 0$ for $0 \leq t \leq T-1$, which we found works best.} 
\stress{Action subspaces:}
Consider the following equivalence relation on the trajectory space $\trajs$:
two joint trajectories $\traj, \traj' \in \trajs$ are equivalent if at each time point $t$,
(1) each agent $i$ is on the same lane in $\traj$ as in $\traj'$, and
(2) within each lane, the order of the agents (along the driving direction) is the same in $\traj$ as in $\traj'$.
Now let the subspace collection $(\actionspart_k)_{k \in \partsindices}$ be obtained by taking the (closures of the) resulting equivalence classes.
\footnote{In the two-driver on-ramp scenario of \Cref{fig:archi} and experiments (\Cref{sec:pred}), these subspaces roughly amount to splitting the action space $\actions$ w.r.t.\ (1) time point of merge and (2) which driver goes first. Note that (1) are additional splits beyond the intuitive ones in (2) (see \Cref{fig:archi}), but they help for concavity and for the analysis.}
\thought{WHY IS THIS NEEDED?:, but removing all joint trajectories from each class where two drivers are closer than $\varepsilon > 0$. \todo{how to define this epsilon?}}
\end{scen}

\begin{prop}[\Cref{sett:driving}'s suitability]
\label{prop:driving}
\Cref{sett:driving} satisfies \Cref{asm:diffetc}.
So, in particular, \Cref{thm:implicitlayer}'s implications on the induced implicit layer hold true. 
%
\end{prop}



Our general approach (\Cref{sec:approach}) in principle is also applicable to various other multiagent trajectory settings, such as pedestrian interaction, relevant for mobile robots. 
We analyze a simplistic such scenario in Sec.~\suppref{sec:pedestriandetails}, see \Cref{fig:pedestrians} (right) for a foretaste. 

\section{Experiments}
\label{sec:experiments}

We evaluate our approach on (1) an observational prediction task\footnote{This directly evaluates the method's abilities for the observational/passive prediction task, but it is also a proxy metric/task for decision making.
}
 on two real-world data sets (\Cref{sec:pred}), as well as (2)~a simple decision-making transfer task 
(\Cref{sec:experimentsdecision}).

\subsection{Prediction Task on Highway Merging Scenarios in Two Real-World Data Set}
\label{sec:pred}

%
%

\begin{table*}[t]
	\renewcommand{\arraystretch}{1.1}  
	%
	%
	%
	\begin{tabularx}{\linewidth}{Xccccc} 
		\hlineB{2.5}
		Data set & \quad\quad\quad\quad Metric \quad\quad\quad\quad & TGL (ours) & TGL-D (ours)  & CS-LSTM 
		& MFP 
		\\
		
		\hline 
		\multirow{2}{*}{highD \citep{highDdataset}}  & MAE  & 3.6 & 2.9 & 5.0 & 5.2    \\ \cline{2-6} 
		& RMSE  & 4.9  & 3.7 & 6.8 & 7.1   \\	
		
		\hline 
		\multirow{2}{*}{HEE (our new data set; \Cref{sec:pred})} & MAE   & 3.7 & 3.2 & 3.6 & 3.7     \\ \cline{2-6} 
		& RMSE   & 4.7  & 4.1 & 4.3 & 4.8  \\	
		
		\hlineB{2.5}
	\end{tabularx}
	\caption{\emph{Prediction task:} Our method(s) vs. state-of-the-art (CS-LSTM \citep{deo2018convolutional}, MFP \cite{tang2019multiple}) for a prediction task on merging scenarios in two real-world highway data sets, averaged over a 7s prediction horizon.}
	\label{table:evaldriving}
	\vspace{-.1cm}
\end{table*}


\renewcommand{\pw}{.49\linewidth}
\renewcommand{\ph}{.12\linewidth}
\begin{figure*}[!th]
	\centering
	\includegraphics[width=\pw,height=\ph]{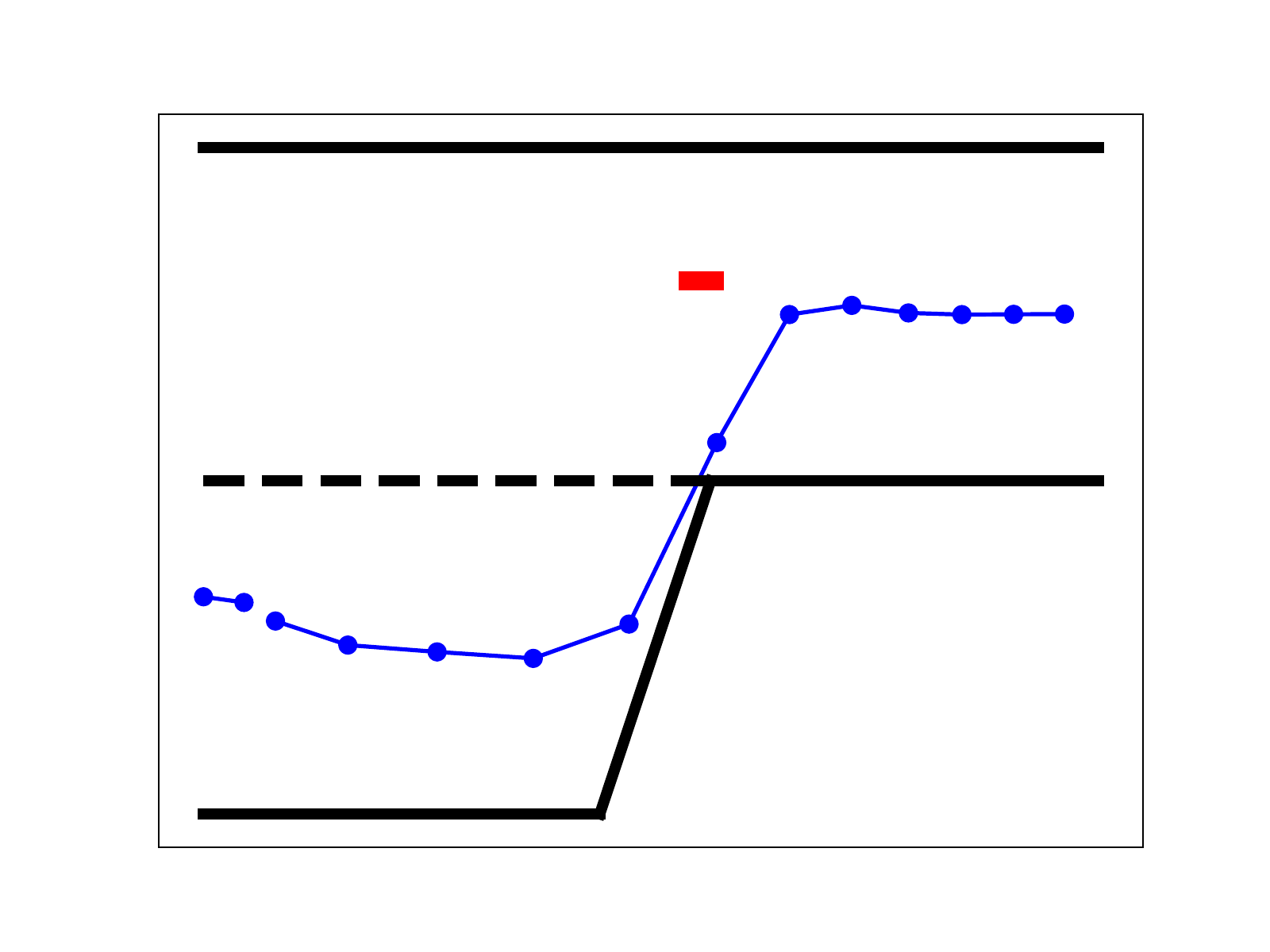}
	\includegraphics[width=\pw,height=\ph]{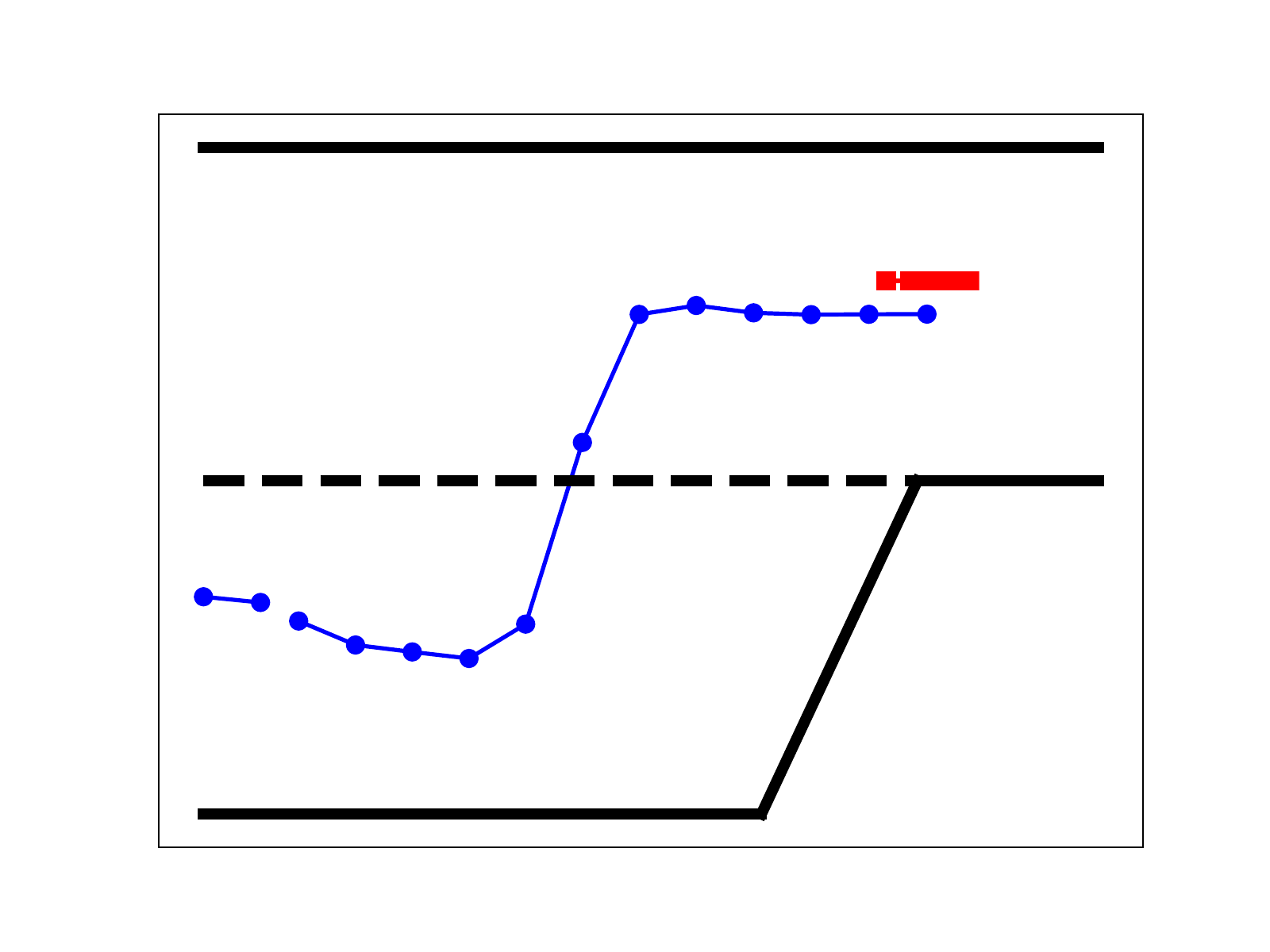}
		\vspace{-.3cm}
	\caption{
		\emph{Decision-making transfer task:} Solution trajectorie(s) that the (partially learned) game implies for the self-driving car's decision-making task (each circle/square corresponds to one time step). \textit{Left:} First local NE: the self-driving car (red) does a full emergency break and the other (blue) merges before it. \textit{Right:} Second local NE: the other merges after it, both slow down.}
	\label{fig:explin}
	\label{fig:self}
\end{figure*}


We consider a highway merging interaction scenario with two cars 
similar as sketched in \Cref{fig:archi}. This is considered a challenging scenario for autonomous driving.

\paragraph{Implementation details for our method for these merging scenario.}
We use the following generic implementation of our general approach (\Cref{sec:approach}), with concrete setting, game and action subspaces from \Cref{sett:driving} (with $\nagents=2$), referring to it as \emph{TGL (trajectory game learner)}: \maybe{We use a smart parameterization of the actions to avoid constraints.}
We use validation-based early stopping.
We combine equilibrium refinement and weighting net into one module, consisting of two nets that
predict the weights $(\flexiscript{\pweight}{k})_{k \in \refpartsindices}$ on the combination of (1) merging order (before/after) probabilities via a cross-entropy loss (2 hidden layers: $1\times 16$, $1\times 4$ neurons; dropout 0.6), and (2) Gaussian distribution over merging time point (discretized and truncated, thus the support inducing a refinement; 2 hidden layers: $1\times 64$, $1\times 32$ neurons; dropout 0.6), given $\inittraj$.
For the preference revelation net we use a feed forward net (two hidden layers: $1\times 16$, $1 \times 24$ neurons).%
\footnote{For varying initial trajectory lengths, an LSTM might be more suitable.}
As training loss we use mean absolute error (MAE; see also evaluation details below).
%
%

Besides this generic instantiation, we also consider a version of it, termed \defi{TGL-D}: Instead of predicting the desired velocity  $\gparam\ai{\nam{v}, i}$ itself, the preference revelation net predicts the difference to the past velocity, and then squashes this into a sensible range using a sigmoid.
(This can be seen as encoding a bit of additional prior knowledge which may not always be easy to specify and depend on the situation.)
For further details, see Sec.~\suppref{sec:expdetails}. 
\urlstyle{tt}
\stress{Code is available at:}
\url{https://github.com/boschresearch/trajectory_games_learning}.


\paragraph{Baselines.}
As baselines we use the state-of-the-art data-driven methods ``convolutional social pooling'' -- specifically: \defi{CS-LSTM} \citep{deo2018convolutional} -- and ``Multiple Futures Prediction'' (MFP) \cite{tang2019multiple}.
\stress{Evaluation.}
We use four-fold cross validation (splitting the data into 4 $\times$ 75\% train and 25\% validation).
As metrics, we use rooted mean squared error (RMSE) and MAE (in meters) between predicted future trajectory $\ptraj$ and truth $\traj$, averaged over a 7s horizon, with prediction step size of 0.2s, applying this to the most likely mode given by our method.


\paragraph{Data sets (one new one) and filtering:}

\emph{1st data set:}
We use the ``highD'' data set \citep{highDdataset}, which consists of car trajectories recorded by drones over several highway sections.
It is increasingly used for benchmarking \citep{rudenko2019human,zhang2020spatiotemporal}. 
From this data set, we use the recordings done over a section with an on-ramp.
%

\emph{2nd data set:}
We publish a \stress{\emph{new data set}} with this paper, termed \stress{\defi{HEE (Highway Eagle Eye)}}. 
It consists of $\sim$12000 individual car trajectories ($\sim$4h), recorded by drones over a highway section (length $\sim$600m) with an entry lane. The link to the data set and further details are in Sec.~\suppref{sec:expdata}.
Keep in mind that this data set can be useful for studies like ours, but some aspects of it may be noisy, so it is only meant for such experimental purposes.

\submission{\footnote{A truncated version of the raw data set and the specific trajectories we use in the experiment are available in the supplement. The full data will be released upon publication. 
		 \old{The specific trajectories we use in the experiment are also contained in a separate supplement file. Compared to the highD data set, the highway section with on-ramp is longer. But highD contains other highway section and is potentially less noisy. (See also the supplement.)}}}
\maybe{\Cref{fig:archi} shows a stylized -- significantly squeezed -- partial picture of the recorded highway section.}

\emph{Selection of merging scenes in both data sets:}
We filter for all joint trajectories of two cars where one is merging from the on-ramp, one is on the rightmost highway lane, and all other cars are far enough to not interact with these two. This leaves 25 trajectories of highD and 23 of our new data set.








\paragraph{Results.}
\old{- bias induced by the simple game may be slightly worse than other biases, more rigid.}
The results are in \Cref{table:evaldriving} (with more details in Sec.~\suppref{sec:expdetails}).
Our generic method TGL outperforms CS-LSTM and MFP on highD. And our slightly more hand-crafted method TGL-D outperforms them on both data set. 
Keep in mind that the data sets are small. So, while the results do indicate the practicality of our method in this small-sample regime, their significance is comparably limited.



\subsection{Simple Decision-Making Transfer Task in Simulation} 
\label{sec:experimentsdecision}

As discussed in \Cref{sec:decisionmaking} the game $\game_\gparam$ -- once $\gparam$ is given, e.g., by our learned preference revelation net -- naturally transfers to decision-making tasks in situations with multiple strategic agents (something which predictive methods like the above CS-LSTM usually cannot do).
To test and illustrate its ability for this, we consider a simple scenario:
Take the above two-car highway on-ramp situation (\Cref{sec:pred}, \Cref{sett:driving}), but assume that the car on the highway lane is a self-driving car.
Assume it has a technical failure roughly at the height of the on-ramp's end, and it should do an emergency break (i.e., desired velocity $\gparam\ai{\nam{v}, i}$ in \eqref{eqn:driver} is set to 0) while at the same time ensuring that the other car coming from the on-ramp will not crash into it.
Which trajectory should it choose?
\stress{Result.} Fed with this situation, our game solver suggests two possible solutions -- two local NE, see \Cref{fig:self}: (1) the self-driving car completely stops and the on-ramp car will merge in front of it, accepting to touch the on-ramp's end; (2) the self-driving car moves slowly, but at a non-zero speed, with the other car right behind it (keeping a rational distance).
While a toy scenario, we feel that these are sensible solutions.
Note that, Additionally, this shows that we (similar to IRL) have a strong ability to reasonably generalize out of sample, since a fully braking vehicle is actually not in the data.

%
%
%
%
%
%


\section{Conclusion}

For modeling of realistic continuous multiagent trajectories, in this work we proposed an end-to-end trainable model class that hybridizes neural nets with game-theoretic reasoning. We accompanied it with theoretical guarantees as well as an empirical demonstration of its practicality, on real-world highway data.
We consider this as one step towards machine learning methods for this task that are more interpretable, verifiable and transferable to decision making. This is particularly relevant for safety-critical domains that involve interaction with humans. 
A major challenge is to make game-theoretic concepts tractable for such settings, and we were only partially able to address this. 
Specifically, potential for future work lies in relaxing subspace-wise concavity, common-coupled games and related assumptions we made.

\section*{Acknowledgments}
We thank Jalal Etesami, Markus Spies and Mathias Buerger for insightful discussions and anonymous reviewers for their hints.


%% file: traffic_games_supp_part.tex
%
%
%

\section{Proofs and remarks} 
\label{sec:proofs}

\subsection{Lemma~\paperref{lem:ispotentialgame}}

Let us first restate the result.


\begin{CLemma}{\paperref{lem:ispotentialgame}}
If $\game_\gparam$ is a common-coupled game, then it is a potential game with the following potential function, where, as usual, $\traj = \parametrization(\action)$:
\[
\potf(\action, \gparam) = \int_0^\finaltime \utilfterm\ai{\shortcommon, \gparam}_t(\traj_{t}) + \sum_{i \in \agents} \utilfterm\ai{\shortown, i, \gparam}_t(\traj^i_{t}) d \mu(t) .
\] 
\end{CLemma}

\begin{proof}[Proof of Lemma~\paperref{lem:ispotentialgame}]
	Recall that based on the definition of a common-coupled game we have have for the stage-wise utility for all $i$ and $t$,
	\begin{align}
	\utilityf\ai{i, \gparam}_t(\traj_{t}) &= \utilfterm\ai{\shortcommon, \gparam}_t(\traj_{t}) + \utilfterm\ai{\shortown, i, \gparam}_t(\traj\ai{i}_{t}) + \utilfterm\ai{\shortothers, i, \gparam}_t(\traj\ai{-i}_{t}).
	\end{align}
	
	Intuitively, the statement directly follows because taking the ``derivative''of $i$'s utility above w.r.t. $\traj\ai{i}$ will just leave the derivative of the common term and $i$'s own term, because the other is constant in $\traj\ai{i}$. And the same happens for $\potf$.

	Formally, observe that for all $i$ and $t$,
	\begin{align*}
	&\utilityf\ai{i, \gparam}_t((\traj\ai[\prime]{i}_{t}, \traj\ai{-i}_{t}))  - \utilityf\ai{i, \gparam}_t((\traj\ai{i}_{t}, \traj\ai{-i}_{t})) \\
	&= \utilfterm\ai{\shortcommon, \gparam}_t((\traj\ai[\prime]{i}_{t}, \traj\ai{-i}_{t})) - \utilfterm\ai{\shortcommon, \gparam}_t((\traj\ai{i}_{t}, \traj\ai{-i}_{t})) \\
	&+ \utilfterm\ai{\shortown, i, \gparam}_t(\traj\ai[\prime]{i}_{t}) - \utilfterm\ai{\shortown, i, \gparam}_t(\traj\ai{i}_{t}) + 0 \\
	&= \utilfterm\ai{\shortcommon, \gparam}_t((\traj\ai[\prime]{i}_{t}, \traj\ai{-i}_{t})) - \utilfterm\ai{\shortcommon, \gparam}_t((\traj\ai{i}_{t}, \traj\ai{-i}_{t})) \\
	&+ \utilfterm\ai{\shortown, i, \gparam}_t(\traj\ai[\prime]{i}_{t}) - \utilfterm\ai{\shortown, i, \gparam}_t(\traj\ai{i}_{t}) + 
	\sum_{j \in \agents \setminus i} \utilfterm\ai{\shortown, j, \gparam}_t(\traj\ai{j}_{t}) - \sum_{j \in \agents \setminus i} \utilfterm\ai{\shortown, j, \gparam}_t(\traj\ai{j}_{t}) .
	\end{align*}
	Integrating w.r.t.\ $t$ and using the linearity of the integral completes the proof.
\end{proof}

\subsection{Theorem~\paperref{thm:implicitlayer}}

Let us first restate the result.

\begin{CTheorem}{\paperref{thm:implicitlayer}}[Game-induced differentiable implicit layer] 
Let \Cpaperref{asm:diffetc} hold true.%
\footnote{Note that this theorem in fact holds for potential games in general, not just common-coupled games. 
	Note also that for tractability/analysis reasons we consider the simple deterministic game form of \Cpaperref{dfn:game} instead of, say, a Markov game.
}
Then, for each $k \in \partsindices$, there is a continuous mapping $\expli_k : \gparams \to \actionspart_k$, such that for any $\gparam \in \gparams$, if $\expli_k(\gparam)$ lies in the interior of $\actionspart_k$, then 

\begin{citem}
	\item $\expli_k(\gparam)$ is a local NE of $\game_\gparam$,
	\item $\expli_k(\gparam)$ is given by the unique argmax\footnote{Due to the concavity assumption, we can use established tractable, guaranteeably sound algorithms to calculate this argmax.} 
	of $\potf (\gparam, \cdot)$ on $\actionspart_k$, with $\potf$ the game's potential function (\Cpaperref{lem:ispotentialgame}), 
	\item  
	$\expli_k$ is continuously differentiable in $\gparam$ with gradient 
	\[\jacobi_\gparam \expli_k (\gparam) =  - \left( \hess_\action \potf (\gparam, \action) \right)^{-1} \jacobi_{\gparam} \nabla_{\action} \potf (\gparam, \action), \] 
	whenever $\potf$ is twice continuously differentiable on an open set containing $(\gparam, \action)$, for $\action = \expli_k (\gparam)$,
	\maybe{\item if $\expli(\gparam)$ lies on (exactly) one constraining hyperplane of $\actionspart_k$ defined by orthogonal vector ... and with strictly positive Lagrange multiplier  ...., then ...}
	where $\nabla, \jacobi$ and $\hess$ denote gradient, Jacobian and Hessian, respectively.
\end{citem}
\maybe{
	In particular, for any $k_1, \ldots, k_\ell$, ............
	\todo{Furthermore, if additionally assumption A2 holds, then ...}
	%
	%
	\begin{citem}
		\item if A1 holds, then, if $\expli(\gparam)$ lies in the interior, then the gradient at $\gparam$ is given by
		\item if A2 holds and if $\expli(\gparam)$ lies in the interior, or if (exactly) the $m$th constraint is active and $\lambda* > 0$, then the gradient
	\end{citem}
	%
	%
	%
}
\end{CTheorem}

\begin{proof}[Proof of Theorem~\paperref{thm:implicitlayer}]

	\cando{The interior-only version. Need to update it for maximum-separating:}	
	
	~\paragraph{Gradient etc.}
	

	Based on Lemma~\paperref{lem:ispotentialgame}, the potential function $\potf$ exists.
	Let $k \in \partsindices$ be arbitrary but fixed.
	Let $\expli_k$ be the function that maps each $\gparam$ to the corresponding unique maximum of $\potf(\gparam, \cdot)$ on $\actionspart_k$ (exists and is unique by the assumption of strict concavity and convexity and compactness of the $\actionspart_k$).
	%
	\old{
		For each $\gparam$, since $\fname{cl}(\actionspart_k)$ is compact, there is a global (on $\fname{cl}(\actionspart_k)$) maximum and thus at least one local maximum (of $\potf (\gparam, \cdot)$).
		At the same time, based on Lemma \paperref{lem:ispotentialgame}, each local maximum is a local NE (of $\game_\gparam$), and so if there were more than one local maxima, then there were more than one local NE, which would contradict our assumption.
		Thus there is exactly one point $\expli_k(\gparam)$ which is a local NE and global maximum, and, by assumption, it is an interior point.
	}
	From the definition of the potential function and the local Nash equilibrium (NE), it follows directly that a maximum of the potential function is a local NE of the game.
	
	To apply the implicit function theorem, let us consider the point $(\gparam, \expli_k(\gparam))$.
	If the minimum $\expli_k(\gparam)$ lies in the interior of $\actionspart_k$, and $\potf$ is continuously differentiable on an open set containing $(\gparam, \expli_k(\gparam))$,
	then we have $\nabla_\action \potf (\gparam, \action) |_{(\gparam, \action) = (\gparam, \expli_k(\gparam))} = 0$ and furthermore, by assumption of strict concavity, $\jacobi_\action \nabla_{\action} \potf (\gparam, \action)  |_{(\gparam, \action) = (\gparam, \expli_k(\gparam))}$ non-singular.
	Then the implicit function theorem implies that there is a an open set $O$ containing $(\gparam, \expli_k(\gparam))$, and a unique continuously differentiable function $f : O \to \actionspart_k$, such that $\nabla_{\action} \potf (\gparam', f(\gparam')) = 0$ for $\gparam' \in O$, with gradient
	\begin{align}
	\jacobi_\gparam f (\gparam) =  - \left( \jacobi_\action \nabla_{\action} \potf (\gparam, \action) \right)^{-1} \jacobi_{\gparam} \nabla_{\action} \potf (\gparam, \action) =  - \left( \hess_\action \potf (\gparam, \action) \right)^{-1} \jacobi_{\gparam} \nabla_{\action} \potf (\gparam, \action).
	\end{align}
	
	Now on $O$, $f$ and $\expli_k$ coincide since $f$ is uniquely determined (specifically, based on the implicit function theorem, locally, the graph of $f$ coincides with the solution set of $\nabla_\action \potf (\cdot, \cdot) = 0$, and if $f, \expli_k$ would differ in at least one point $\gparam'$, then there would be a solution $(\gparam', \expli_k(\gparam'))$ outside the solution set -- a contradiction).
	Therefore $\expli_k$ is also continuously differentiable on $O$ with gradient
	\begin{align}
	\jacobi_\gparam \expli_k (\gparam) = \jacobi_\gparam f (\gparam).
	\end{align}
	
	We can do this for every $\gparam$, which completes the proof.
	
	\cando{Slightly open Qs:
		- due to the current formulation where we can identify the one critical point that matters, it doesnt matter that there could be several critical points, right?
	}

	~\paragraph{Continuity.}
	Since $\potf$ is continuous, $\actionspart_k$ is compact, and the maxima are unique, the \socalled{maximum theorem} implies that the mapping $\expli_k$ is in fact continuous (hemicontinuity reduces to continuity when the correspondence is in fact a function).
	
	\reasonwhy{a main reason why i wanted all this unique optimum/stationary point reasoning was that one possible argument that the implicit mapping is (1) well defined (a uinique ouput) but also (2) have a guaranteed search procedure for it. for both, 1 and 2, there can be other arguments! but maybe we can weaken 2 while keeping 1 for now.}
	
	
\end{proof}

\maybe{
\subsection{\todo{HJB observation}}
\todo{We do not write a rigorous theorem/lemma here because HJB etc. requires some conditions ... convexity .... this is more of a direction in which to go to ...}
\[\utilityf\ai{i}(\action) = \int_0^\finaltime \utilityf\ai{i}_t(\traj_{[t - \Delta, t]}) d \mu(t), \]
\begin{align}
\potf() = \int_0^S  + \int_S^T 
\end{align}
}

\subsection{Lemma~\paperref{lem:gradbound} with remark on zero gradient}

Let us first restate the result.

\begin{CLemma}{\paperref{lem:gradbound}}
	Let \asmref{asm:diffetc} hold true and additionally assume $\potf$ to be continuously differentiable on (a neighborhood of) $\gparams \times \actionspart_k, k \in \partsindices$. 
	If $a=\expli_k(\gparam)$ lies on (exactly) one constraining affine\old{This ``affine'' was missing in the main text.} hyperplane of $\actionspart_k$, defined by orthogonal vector $\constr$, with multiplier $\lambda$ and optimum $\lambda^* > 0$ of $\potf(\gparam, \action)$'s Lagrangian (details in the proof), then
	$
	\jacobi_{\gparam} \expli_k (\gparam) =  
	\left[
	-
	\left(
	\begin{array}{cc}
	\hess_a \potf(\gparam, \action)  & \constr \\
	\lambda^* \constr^T & 0 \\
	\end{array}
	\right)^{-1} 
	\left(
	\begin{array}{c}
	\jacobi_{\gparam} \nabla_\action \potf(\gparam, \action) \\
	0 
	\end{array}
	\right) \right]_{1:(\nagents \cdot\adim) \times 1:(\nagents \cdot\adim)}
	$.
\end{CLemma}

\paragraph{Remark on zero gradient.}
Note that under the conditions of this lemma, i.e., when $\expli_k (\gparam)$ lies at the boundary, then the above gradient $\jacobi_{\gparam} \expli_k (\gparam)$ in fact often becomes zero, which can be a problem for paramter fitting.
So the above result is only meant as a first step.

\begin{proof}[Proof of Lemma~\paperref{lem:gradbound}]



	\newcommand{\multio}{\multi^*}
	\newcommand{\opti}{\lne}
	\newcommand{\bound}{b}
	
	
	
	\cando{calculate an example to check that the precise gradient formula is correct}
	
	Let $\actionspart_k$ be defined by the inequality constraints $\constr_m^T \action \leq \bound, m=1, \ldots, M$.
	Consider the Lagrangian
	\begin{align*}
	\Lambda(\gparam, \action, \multi_1, \ldots, \multi_M) = \potf(\gparam, \action) + \sum_m \multi_m ( \constr_m^T \action - \bound).
	\end{align*}
	
	Then the Karush-Kuhn-Tucker optimality conditions \cite{boyd2004convex} (note that we assumed differentiability of $\potf$ on a a neighborhood of $\gparams \times \actionspart_k, k \in \partsindices$) include the following equations:
	\begin{subequations}
		\label{eqn:kkt}
		\begin{align}
		\nabla_a \Lambda(\gparam, \action, \multi_1, \ldots, \multi_M) &= 0, \\
		\multi_m (\constr_m^T \action - \bound ) &= 0,  \ m=1, \ldots, M.
		\end{align}
	\end{subequations}
	
	Now
	let $\opti = \expli_k(\gparam)$ and let $\multio_1, \ldots, \multio_M$ be the (optimal) duals for $\opti$.
	And assume that $\opti$ lies on exactly one bounding (affine) hyperplane.
	W.l.o.g. let this hyperplane correspond to $\constr_1, \multio_1$.
	Also recall that $\expli_k$ is continuous (as in Theorem \paperref{thm:implicitlayer}).
	Therefore, in a neighborhood of $\gparam$, the corresponding optimum will not lie within any of the other boundaries.
	So in this neighborhood of $\gparam$, all corresponding optimal duals will be zero (inactive) except for the assumed one.
	
	Therefore, given a $\gparam'$ from the mentioned neighborhood of $\gparam$, we have that $\action, \constr_1$ satisfy the optimality conditions in \eqref{eqn:kkt} (for some remaining duals) iff they satisfy the reduced conditions
	\begin{subequations}
		\label{eqn:kkt2}
		\begin{align}
		\nabla_a \Lambda(\gparam', \action, \multi_1, \multi, 0, \ldots, 0) &= 0, \\
		\multi_1 (\constr_1^T \action - \bound ) &= 0.
		\end{align}
	\end{subequations} 
	
	%
	%
	For succinctness, in what follows we write $\multi$ instead of $\multi_1$, i.e., drop the subscript.
	%
	%
	%
	%
	Let
	\begin{align*}
	h(\gparam', \action, \lambda) 
	&:= ( \nabla_{\action} \Lambda(\gparam', \action, \lambda, 0, \ldots, 0), \lambda (\constr^T \action - \bound ) ) \\
	&= ( \nabla_{\action} \potf(\gparam', \action) + \nabla_{\action} \lambda ( \constr^T \action - \bound), \lambda (\constr^T \action - \bound ) )\\
	&= ( \nabla_{\action} \potf(\gparam', \action) + \lambda \constr^T, \lambda (\constr^T \action - \bound ) )
	\end{align*}
	So the conditions in \eqref{eqn:kkt2} are
	\begin{align}
	\label{eqn:h}
	h(\gparam', \action, \multi) = 0.
	\end{align}

	
	Similar as in Theorem \paperref{thm:implicitlayer}, around the point $\gparam, \opti, \multio$, $\expli_k$ satisfies \eqref{eqn:h} (for some $\lambda$'s). So we can apply the implicit function theorem to get its gradient.

	%
	%
	
	We have
	\begin{align*}
	&\jacobi_{(\gparam, \action, \lambda)} h(\gparam, \action, \lambda) \\
	&=  \left(
	\begin{array}{ccc}
	\jacobi_{\gparam} \potf(\gparam, \action) & \hess_a \potf(\gparam, \action)  & \constr^T \\
	0 & \lambda \constr^T & \constr^T \action - \bound \\
	\end{array}
	\right) .
	\end{align*}
	
	Note that 
	\begin{align*}
	\left(
	\begin{array}{cc}
	\hess_a \potf(\gparam, \opti)  & \constr^T \\
	\multio \constr^T & \constr^T \opti - \bound \\
	\end{array}
	\right)
	=
	\left(
	\begin{array}{cc}
	\hess_a \potf(\gparam, \opti)  & \constr^T \\
	\multio \constr^T & 0 \\
	\end{array}
	\right)
	\end{align*}
	is invertible, since 
	\begin{align*}
	\det\left( \left(
	\begin{array}{cc}
	\hess_a \potf(\gparam, \opti)  & \constr^T \\
	\multio \constr^T & 0 \\
	\end{array}
	\right) \right) = \det( \hess_a \potf(\gparam, \opti) ) \det(- \multio \constr^T (\hess_a \potf(\gparam, \opti))^{-1} \constr )
	\end{align*}
	and both factors are non-zero since $\hess_a \potf(\gparam, \opti)$ is positive definite and $\multio, \constr$ are non-zero.
	
	Therefore, the implicit function theorem is applicable to the equation, and we get as gradient
	\begin{align*}
	\jacobi_{\gparam} \expli_k (\gparam) =  -
	\left(
	\begin{array}{cc}
	\hess_a \potf(\gparam, \opti)  & \constr^T \\
	\multio \constr^T & 0 \\
	\end{array}
	\right)^{-1} 
	\left(
	\begin{array}{c}
	\jacobi_{\gparam} \potf(\gparam, \opti) \\
	0 
	\end{array}
	\right) 
	\end{align*}

\end{proof}

\subsection{Proposition \paperref{prop:driving}}

Let us first restate the result.

\begin{CProposition}{\paperref{prop:driving}}[\Cref{sett:driving}'s suitability]
	\Cref{sett:driving} satisfies \Cref{asm:diffetc}.
	So, in particular, \Cref{thm:implicitlayer}'s implications on the induced implicit layer hold true. 
	%
\end{CProposition}

\newcommand{\argmax}{\nam{arg max}}
\newcommand{\argmin}{\nam{arg min}}
\newcommand{\subsp}{\tilde{\actions}}

\begin{proof}[Proof of Proposition \paperref{prop:driving}]
	
	\thought{asm that i found out were necessary: the multiplicative parameters of velocity and the acceleration term have to be non-zero (otherwise still convex but not strictly). 
		btw: is it enough to have one term that in the full vector is strictly convex and then arbitrarily many that are potentially weakly convex? i guess so, based on additive pd argument}

	\thought{also need to show differentiaability?}

	~\paragraph{Common-coupled game.}
	It is directly clear from the form of the utilities in Scenario \paperref{sett:driving}, that this forms a common-coupled game.

	~\paragraph{Strict concavity of the potential function.} 
	
	
	Observe that, for each $i$, within any one subspace $\actionspart_k$, for each $t$, 
	\begin{itemize}
		\item $i$ does not changes lane, so $c_t, e_t$ are simple constants.
		\item For each lane, there is a fixed set of agents. Consider the set $S$ of pairs $(j, j')$ of agents that are on this lane and $j$ is right before $j'$. This ordering (and thus $S$) is invariant within $\actionspart_k$. Therefore the agent distance term can be rewritten like
		\begin{align}
		&\gparam\ai{\nam{dist}} \sum_{\text{$j$ right before $j'$ on same lane}}\frac{1}{ | \posx\ai{j}_t - \posx\ai{j'}_t | + \zeta} \\
		&= \gparam\ai{\nam{dist}} \sum_{{(j, j') \in S}}\frac{1}{  \posx\ai{j'}_t - \posx\ai{j}_t  + \zeta}
		\end{align}
	\end{itemize}

	So all terms are concave (the other terms are obviously concave), therefore the overall potential function, which is just a sum of them, is concave.
	
	Futhermore, note that the sum of all velocity and distance to lane center terms is a sum of functions such that for each component of the vector $\action$ there is exactly one function of it, and only of it; and each function is strictly concave. This implies that the overall sum is strictly concave in the whole $\action$. So the potential function is a sum of concave and a strictly concave term, meaning it is strictly concave.

	NB: On the subspaces, the potential function is also differentiable.

	%
	%

	\paragraph{Compactness and linearity of constraints.}
	
	Besides the constraints that define the compact complete action space $\actions$, which are obviously linear,
	the constraints that define the action subspaces $\subsp_k$ are given by the intersection of constraints for each time point $t$ that are all of the form
	\begin{itemize}
		\item $\posy\ai{i}_t \geq \nam{const.}$ or $\posy\ai{i}_t \leq \nam{const.}$, or
		\item $\posx\ai{i}_t \leq \posx\ai{j}_t$,
	\end{itemize}
	so they are linear.

\end{proof}

\section{Further details on general architecture}
\label{sec:tgldetails}

Elaborating on Sec.~\paperref{sec:archi}: 
\Cref{alg:tglf} describes TGL's forward pass, in particular the parallel local NE search over the refined subspaces.
\Cref{alg:tglt} describes the training more formally, in particular the splitting into first training the equilibrium refinement net alone.

\begin{algorithm}[t]
	\caption{TGL -- forward pass (sketch)}
	\label{alg:tglf}
	\KwIn{past joint trajectory $\inittraj$}
	\KwOut{modes $\flexiscript{\ptraj}{k}$ and their probabilities $\flexiscript{\pweight}{k}$, $k \in \refpartsindices$, as prediction for future joint trajectory $\traj$		} 
	$\gparam$ $=$ preference revelation net $(\inittraj)$\footnotemark{} \;
	$\refpartsindices$ $=$ equilibrium refinement net $(\inittraj)$\; 
	\tcp{Implicit layer $\expli$:} 
	\ForEach{$k \in \refpartsindices$ \label{line:f}}{
		\tcp{Solve for local NE $\flexiscript{\lne}{k}$ in $\actionspart_k$:}
		$\flexiscript{\lne}{k} = \expli_k(\gparam) = \arg\max_\action \potf (\gparam, \action)$ s.t. ${\action \in \actionspart_k}$%
		\label{line:fe}%
		\;
	}
	$(\flexiscript{\ptraj}{k})_{k \in \refpartsindices}$ $=$ trajectory parametrization $\parametrization$ $((\flexiscript{\lne}{k})_{k \in \refpartsindices})$ \; 
	$(\flexiscript{\pweight}{k})_{k \in \refpartsindices}$ $=$ eq.\ weighting net $( 
	\gparam, (\flexiscript{\lne}{k})_{k \in \refpartsindices}),  \ldots)$\;
\end{algorithm}

 \addtocounter{footnote}{-1} 
\stepcounter{footnote}\footnotetext{
	``A net (B)'' reads ``output of the A net applied to input B''.}

\begin{algorithm}[!tbp]
	\caption{TGL -- training (sketch, in particular how to separate the equilibrium refinement net)}
	\label{alg:tglt}
	\KwIn{set $\trset$ of training samples $(\inittraj', \traj'), (\inittraj'', \traj''), \ldots$}
	\tcp{Train equilibrium refinement net:}
	Train the weights $w_{\nam{er}}$ of the equilibrium refinement net on pairs $(\inittraj, k)$, where $k \in \partsindices$ is the index of the subspace the ground truth $\traj$ lies in, for $(\inittraj, \traj) \in \trset$ \;
	\tcp{Train full architecture but with the equilibrium refinement net's weights fixed:}
	Perform gradient-based training of the full architecture by training all its weights except for the  equilibrium refinement net's weights $w_{\nam{er}}$ which are fixed to the one obtained by the training above. The forward pass is given by \Cref{alg:tglf}. To get gradients via back-propagation, for the implicit layer's gradient use the formula given by \Cref{thm:implicitlayer} \;\old{.... multiplied with the other layers gradients in the usual chain rule ...}
\end{algorithm}

\section{Further example scenario: simple pedestrian encounter}
\label{sec:pedestriandetails}

\begin{wrapfigure}{l}{.2\linewidth}  
	\includegraphics[width=\linewidth]{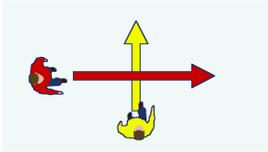}
	\caption{Simplistic pedestrian encounter.} 
	\label{fig:pedestriansd}
\end{wrapfigure}

As indicated in the main text, let us now elaborate the second -- pedestrian scenario -- our general method applies to.
When considering settings with characteristics such as continuous time, even if they still satisfy the conditions of our framework, to \emph{prove} so can become arbitrarily complex. 
Here let us give a simplistic but verified second example with properties somewhat different from the first one:
%
Consider two pedestrians who walk with constant velocity (this velocity is their respective action) along straight paths which are orthogonal and intersect such that they could bump into each other (\Cref{fig:pedestriansd}).
Formally:

\begin{scen}[Simple pedestrian encounter]
	\label{sett:pedestrian}
	\stress{Setting:} There are
	$\nagents=2$ agents,
	the actions parameterize the trajectories\old{\footnote{\todo{CAN DROP(?): To be rigorous, similar as in \Cref{sett:driving}, actually we have to augment the state at each time $t$, e.g., with $\max_{t' \in [0, t]} \frac{1}{\| \traj\ai{1}_{t'} - \traj\ai{2}_{t'} \|_1}$.}}} via $\traj\ai{1}_t = (0, t \action\ai{1} + z\ai{1})$, $\traj\ai{2}_t = (t \action\ai{2} + z\ai{2}, 0)$ (for ease of notation, we put the intersection to the origin but translations are possible of course),
	the joint action space is $\actions = [\frac{z\ai{1}}{\finaltime}, c] \times [\frac{z\ai{2}}{\finaltime}, c]$,
	for constants $z\ai{1}, z\ai{2} < 0$ and $c > 0$,
	(the lower bound on the action is to make sure they reach the intersection).
	\stress{Game:} \maybe{To allow to take maxima over the trajectory and still fit into our general setting (\Cref{sec:generalsetting}),} 
	Let the final stage utility be given by the following sum of a distance penalty term and a desired velocity term
	\begin{subequations}
		\begin{align}
		&\utilfterm\ai{i, \gparam}_\finaltime(\traj^i_{T})
		= - h (\gparam\ai{\nam{dist}}, \action) \max_{t \in [0, \finaltime]} \frac{1}{\| \traj\ai{1}_t - \traj\ai{2}_t \|_1} \\ 
		&- \gparam\ai{\nam{vel}, i} (\action\ai{i} - \gparam\ai{\nam{v}, i}     )^2 \text{ ($=-\infty$ if division by 0) }, 
		\end{align}
	\end{subequations}
	for some function $h$,
	$ \gparam\ai{\nam{vel}, i} > 0$,
	and let $\mu$ be the Dirac measure on $\finaltime$.
	\old{, and $\Delta = \finaltime$
	(We use this terminal-term and Dirac delta-based formulation to properly fit it into the general \Cref{dfn:game}).}
	\stress{Action subspaces:} Let the subspaces $(\actionspart_k)_{k \in \partsindices}$ be given by (1) taking the two subspaces that satisfy $\action_k \geq \action_{2-k} + \varepsilon$, for $k=1,2$ respectively, and some small $\varepsilon > 0$, i.e., split by which agent is faster, and (2) additionally split by which agent first reaches their paths' intersection (altogether this yields two or three subspaces). 
	\maybe{drop second split that we dont require differentiability everywhere anymore?}
	\maybe{e.g. for the merging scenario .... fig 1 ... we separate by time point of lane change and whether red or yellow goes first.}
\end{scen}


\begin{prop}[\Cref{sett:pedestrian}'s suitability and partial identifiability]
	\label{prop:pedestrian}
	Assume \Cref{sett:pedestrian} with $h(\gparam\ai{\nam{dist}}, \action) = \frac{1}{\action_k}$, with $k$ the faster agent in $\action$. Then \Cref{asm:diffetc} is satisfied.
	\old{Then the conditions of \Cref{thm:implicitlayer} are satisfied.\footnote{Due to taking a maximum over the full trajectory,
			this scenario does not exactly fit \Cref{asm:diffetc}, but a version of it, see proof.}}
	Furthermore, while the complete game parameter vector $\gparam = (\gparam\ai{\nam{vel}, 1}, \gparam\ai{\nam{v}, 1}, \gparam\ai{\nam{vel}, 2}, \gparam\ai{\nam{v}, 2}, )$ is not identifiable in general, if $\gparam\ai{\nam{vel}, i}, i=1,2$ is constant, then $(\gparam\ai{\nam{v}, 1} ,\gparam\ai{\nam{v}, 2} )$ is identifiable from $\traj$ on the pre-image of the interior of $\actionspart_k$, for any $k$.
\end{prop}

\subsection{Proof of Proposition \ref{prop:pedestrian}}

%

\begin{proof}[Proof of Proposition \paperref{prop:pedestrian}]
	
	\newcommand{\ic}{z}
	\newcommand{\coord}{z}
	
	~\paragraph{Remark on utility form.}
	Note that in Scenario \paperref{sett:pedestrian} the stage-utility at time $T$ in its ``distance term'' uses a form of maximum operator that takes the whole trajectories as input. But according to our Definition \paperref{dfn:game}, actually we allow the stage utility to only depend on respective current state.
	But this can be resolved by observing that the ``distance term'' can in fact be rewritten as a function of $(\action\ai{1}, \action\ai{2})$, as we will see in \Cref{eqn:a20} below.
	And $(\action\ai{1}, \action\ai{2})$ in turn is given by $(\traj\ai{\finaltime}_1 - \ic\ai{1}, \traj\ai{\finaltime}_2 - \ic\ai{2}) / \finaltime$, i.e., determined by $\traj\ai{\finaltime}$.

	\paragraph{Common-coupled game.}
	It is directly clear from the form of the utilities in Scenario \paperref{sett:pedestrian}, that this forms a common-coupled game.
	
	\paragraph{Strict concavity of the potential function.}
	


	Consider all subspaces $\actionspart_k$ where agent 1 is faster, i.e., $\action_1 > \action_2$. 
	Note that this can be one or two subspaces: if $\traj\ai{0}_1 > \traj\ai{0}_2$, then it is one subspace, but otherwise it is two (namely, where 1 or 2 reaches the intersection first, respectively).
	
	On any one of these subspaces, the following holds true:
	

	Regarding the distance-based term, observe that 
	\begin{align}
	\argmin_{t \in [0, \finaltime]} | t \cdot \action\ai{1} + \ic\ai{1} | + | t \cdot \action\ai{2} + \ic\ai{2} | 
	= - \frac{\ic\ai{1}}{\action\ai{1}} . 
	\end{align}
	
	To see this, observe that the argmin is given by $t$ where agent 1 reaches the intersection, i.e, $t$ where $t \cdot \action\ai{1} + \ic\ai{1} = 0$, i.e., $t = - \frac{\ic\ai{1}}{\action\ai{1}}$. (And this holds regardless of whether 1 or 2 first reaches the intersection.)
	
	To see this in turn, first consider the case that agent $1$ first reaches the intersection.
	Then, before this $t$, both terms are bigger (both agents are further away from the origin),
	while at this $t$, the first term is 0, and after it the left term grows faster (because the agent is faster) than the right term decreases (until the right term hits $0$ as well and then also increases again).
	
	Second, consider the case where agent $2$ first reaches the intersection (in case this happens at all -- i.e., if 2 starts so much closer to the origin to make this possible).
	Then, before $2$ reaches the intersection: obviously both terms are bigger than when $2$ reaches the intersection.
	Between $2$ reaching the intersection and $1$ reaching the intersection: in this time span, the right term grows slower than the left term decreases, therefore the minimum (for this time span) happens when 1 reaches the intersection.
	Now after 1 has reached the intersection obviously both terms just grow.

	Therefore,
	\begin{align}
	\utilfterm\ai{\shortcommon, \gparam}_\finaltime(\traj_{[0, \finaltime]}) 
	&= - \frac{1}{\action\ai{1}} \max_{t \in [0, \finaltime]} \frac{1}{\| (0, t \cdot \action\ai{1} + \ic\ai{1}) - (t \cdot \action\ai{2} + \ic\ai{2}, 0) \|_1 } \\
	&= - \frac{1}{\action\ai{1}} \max_{t \in [0, \finaltime]} \frac{1}{| t \cdot \action\ai{1} + \ic\ai{1} | + | t \cdot \action\ai{2} + \ic\ai{2} |} \\
	&= - \frac{1}{\action\ai{1}} \frac{1}{| - \frac{\ic\ai{1}}{\action\ai{1}} \cdot \action\ai{1} + \ic\ai{1} | + | - \frac{\ic\ai{1}}{\action\ai{1}} \cdot \action\ai{2} + \ic\ai{2} |} \\
	&= - \frac{1}{\action\ai{1}} \frac{1}{| - \frac{\ic\ai{1}}{\action\ai{1}} \cdot \action\ai{2} + \ic\ai{2} |} \\
	&= - \frac{1}{| \ic\ai{2} \action\ai{1} - \ic\ai{1} \action\ai{2}  |} \label{eqn:a20}
	\end{align}
	
	Keep in mind that for agent $i$, the time when it reaches the intersection is given by $t_i = - \frac{\ic\ai{i}}{\action\ai{i}}$.
	Now, if, for the subspace under consideration, agent 1 first reaches the intersection, i.e., 
	$- \frac{\ic\ai{1}}{\action\ai{1}} > - \frac{\ic\ai{2}}{\action\ai{2}}$, i.e.,
	$\frac{\ic\ai{1}}{\action\ai{1}} < \frac{\ic\ai{2}}{\action\ai{2}}$, i.e.,
	$\ic\ai{1} \action\ai{2} < \ic\ai{2} \action\ai{1}$.
	Then \eqref{eqn:a20} becomes $- \frac{1}{\ic\ai{2} \action\ai{1} - \ic\ai{1} \action\ai{2}}$,
	which is obviously concave.
	Similarly for the case that agent 2 first reaches the intersection.
	
	Regarding the velocity terms, obviously their sum is strictly concave.
	
	So the sum of all terms is strictly concave.
	

	\paragraph{Linearity and compactness of constraints.}
	The constraints are all of the form $\action_1 \geq \action_2 + \varepsilon$ or $\ic\ai{1} \action\ai{2} < \ic\ai{2} \action\ai{1}$, i.e., linear.

	\paragraph{Identifiability.} 
	
	Obviously the full $\gparam$ cannot be identifiable because there are no (local) diffeomorphisms between spaces of differing dimension.

	\newcommand{\inte}{\fname{int}}
	
	
	
	Keep in mind that the parametrization from $\action$ to $\traj$ is injective, so we just need to show identifiability from $\action$.
	That is, we have to show that $\expli_k$ is invertible on $\expli_k^{-1}(\inte(\bigcup_k \actionspart_k))$, where $\inte(\cdot)$ denotes the interior. 
	\cando{(Given $\action$, we know which of the equilibrium-separating subspaces $\actionspart_k$ it lies in; only consider this subspace for now.)} \cando{(because given $\action$, of course we can tell which subspace it lies in?!?! maybe not if they intersect at the boundary?!)}
	
	Since we fixed $\gparam\ai{\nam{vel}, i}, i=1,2$, consider them as constants, and for what follows, for simplicity
	let $\gparam$ stand for $(\gparam\ai{\nam{v}, 1} ,\gparam\ai{\nam{v}, 2} )$.

	W.l.o.g. (the other cases work similarly) assume $\actionspart_k$ is the subspace where agent 1 is faster and first reaches the intersection, so the potential function becomes
	\begin{align}
	\potf (\gparam, \action) = \frac{1}{\ic\ai{2} \action\ai{1} - \ic\ai{1} \action\ai{2}} 
	- \gparam\ai{\nam{vel}, 1} (\action\ai{1} - \gparam\ai{\nam{v}, 1}     )^2
	- \gparam\ai{\nam{vel}, 2} (\action\ai{2} - \gparam\ai{\nam{v}, 2}     )^2 .
	\end{align}
	
	Now let $\action$ be such that $\action = \expli_k(\gparam)$ for some $\gparam$.
	We have to show that there can only be one such $\gparam$.
	To see this, note that $\action$ is a local NE, and thus
	\begin{align}
	0  = \nabla_\action \potf (\gparam, \action) = \left( (-1)^{i-1} \frac{\ic\ai{3-i}}{ (\ic\ai{2} \action\ai{1} - \ic\ai{1} \action\ai{2})^2 } +   2 \gparam\ai{\nam{vel}, i} \action\ai{i} - 2 \gparam\ai{\nam{vel}, i} \gparam\ai{\nam{v}, i} \right)_{i=1,2} .
	\end{align}
	But this implies
	\begin{align}
	\gparam\ai{\nam{v}, i} = \frac{(-1)^{i-1} \frac{\ic\ai{3-i}}{ (\ic\ai{2} \action\ai{1} - \ic\ai{1} \action\ai{2})^2 } - 2 \gparam\ai{\nam{vel}, i} \action\ai{i}}{2 \gparam\ai{\nam{vel}, i} } .
	\end{align}

\end{proof}

\section{Further details on experiments and implementation}
\label{sec:expdetails}


Keep in mind that in the main text (\Cpaperref{sec:experiments}), we already described two implementations/versions we propose as instantiations of our general method (\Cpaperref{sec:archi})  for the highway merging setting:
\begin{itemize}
	\item TGL,
	\item TGL-D.
\end{itemize}

Here we present one further method:
\begin{itemize}
	\item TGL-DP (also ``TGL-P'' for short) -- building on TGL-D, we use the interpretable representation of the desired velocity parameter predicted by the preference revelation net, which we first validate, and then encode additional \emph{prior-knowledge based constraints} (e.g., we clip maximum and minimum desired speed) --  see \Cref{sec:tglp} for details.
\end{itemize}

Additionally, in this section we provide further details on the new data set (\Cref{sec:expdata}) as  well as more fine-grained empirical evaluations metrics (individual prediction time points) for all methods.

\cando{R1 details about experimental setup, particularly that from R3, should be addressed in the appendix.}

\cando{R3 In section 5.1, how do you justify the shape of the networks you use? Why does a network have only 1 layer, and the second layer has 2? You talk about Gaussian distribution over merging time point. Distribution of what?
	What corresponds, how is used and where is defined the "Merging time point". Is it related to "averaged over a 7s prediction horizon" which is mentioned in the legend of table 1.
	What do the numbers in table 1 represent? In what unit are they given? I imagine that your goal is not to gain in computing speed (otherwise, we have nothing on the type of machine used) but surely in quality (and therefore, how is it measured). Table 1 says that you get 3.7 instead of 5.2 ? so what if it's not ?
	The paragraph "Results" just before 5.2 is very difficult to understand: "Our general method TGL outperforms ..." how does it outperform?}

\begin{figure*}[!htbp]
	\centering
	\includegraphics[width=\linewidth]{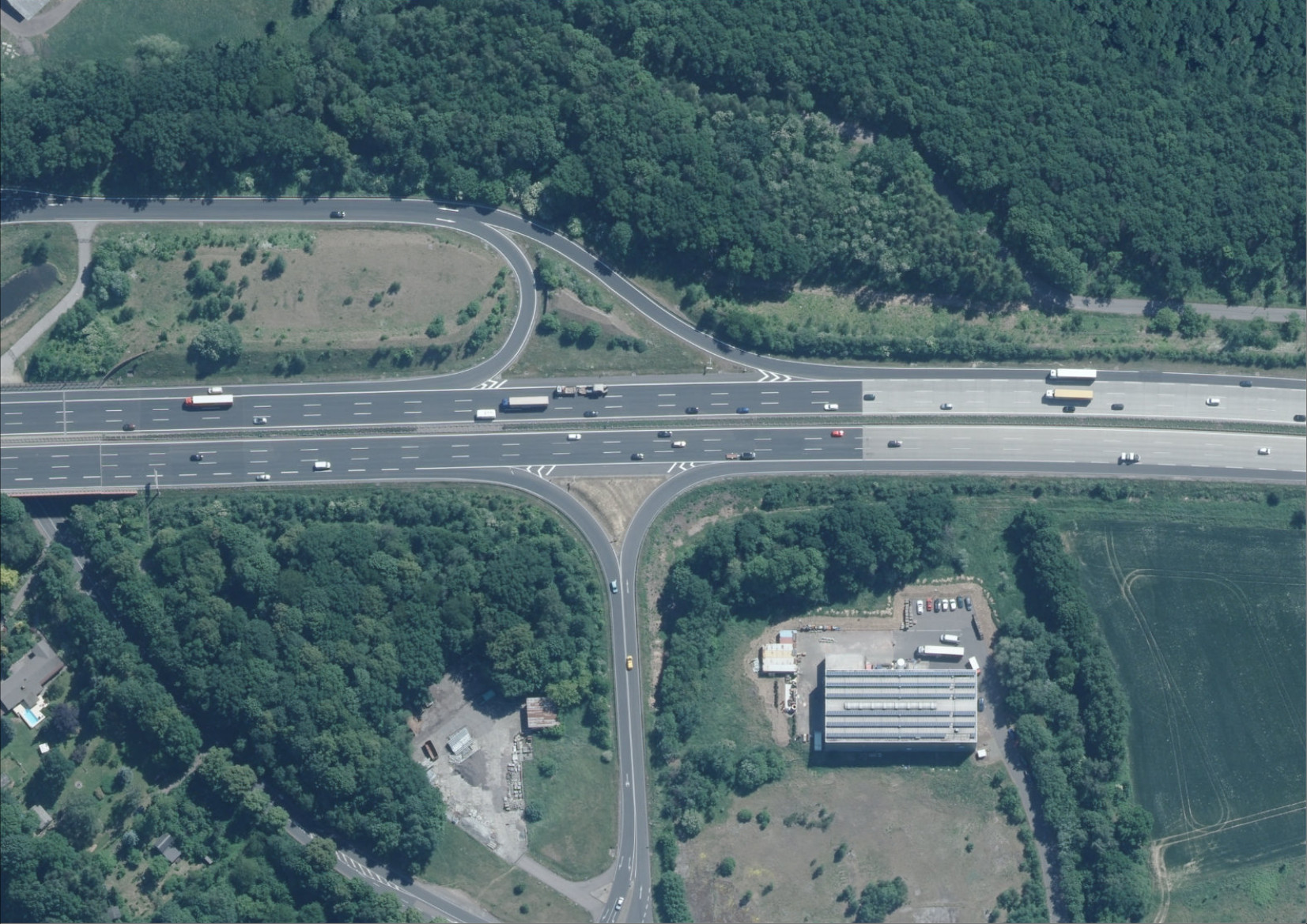}
	\caption{For the new highway data set \submission{published alongside this paper}, this is roughly the recorded highway section (only the lower lane, incl.\ exit/entry). Note that the recorded section is in fact slightly more to the right than the picture indicates.}
	\label{fig:heefull}
	%
	\vspace{.5cm}
	\includegraphics[width=\linewidth]{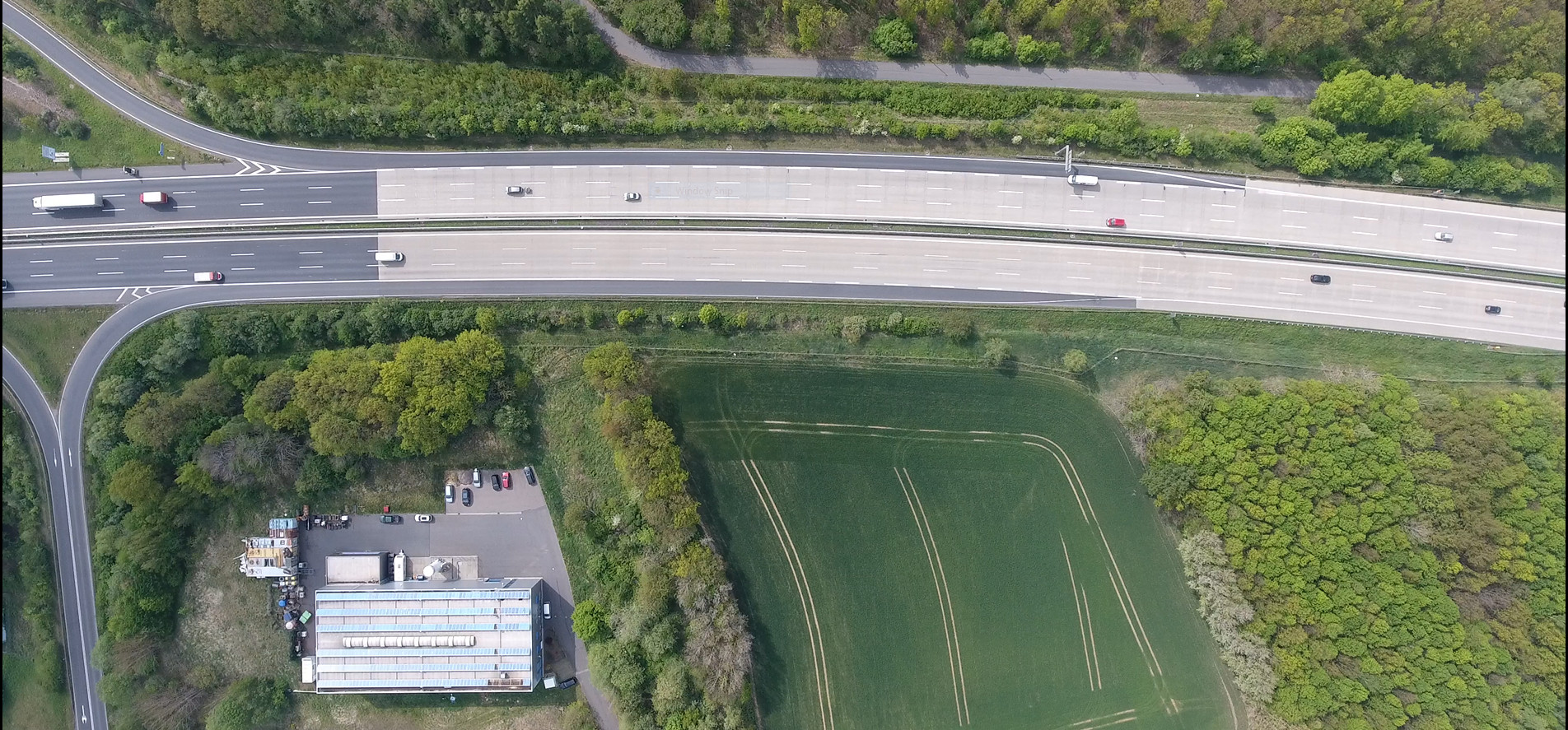}
	\caption{This is a zoom-in on roughly the sub-part of the highway section in \Cref{fig:heefull} that is relevant for the on-ramp merging trajectories used in the experiment.}
	\label{fig:heemerge}
\end{figure*}

\newcommand\mycommfont[1]{\footnotesize\ttfamily\textcolor{darkgray}{#1}}
\SetCommentSty{mycommfont}

\subsection{More detailed description of TGL-DP}
\label{sec:tglp}

\begin{algorithm}[t]
	\caption{Part of preference revelation net of TGL-P that outputs the desired velocity game parameters $\gparam\ai{\nam{v}, 2}, \gparam\ai{\nam{v}, 1}$ and differs compared to TGL}
	\label{alg:tglp}
	\KwIn{
		old\_vx\_other,old\_vx\_merger  \tcp{velocities along x-axis at the last step of the past trajectory $\inittraj$, where ``merger'' is the on-ramp car, and ``other'' is the highway car}\
		desired\_vx\_other,desired\_vx\_merger \tcp{output $\gparam\ai{\nam{v}, 2}, \gparam\ai{\nam{v}, 1}$ of TGL's original preference revelation net}\
		merger\_in\_front \tcp{most probable subspace given by equilibrium refinement/weighting net, whether merger merges before or after other}\
		big\_change \tcp{parameter -- factor for allowed big change, for the experiment we used big\_change = 1.2 }\
		small\_change \tcp{parameter -- factor for allowed small change, for the experiment we used small\_change = 1.04 }
	} 
	\KwOut{
		new\_vx\_other,new\_vx\_merger  \tcp{new desired velocity game parameters $\gparam\ai{\nam{v}, 2}, \gparam\ai{\nam{v}, 1}$}
	} 
	
	\If{merger\_in\_front == 0}{
		\tcp{if the highway vehicle is in front}
		\tcp{clamp change of highway vehicle desired speed} 
		new\_vx\_other = clamp(desired\_vx\_other, old\_vx\_other / small\_change, old\_vx\_other*small\_change)	\\
		\tcp{only allow accelerating merger vehicle, but limit maximum speed change}
		new\_vx\_merger = clamp(desired\_vx\_merger, old\_vx\_merger, old\_vx\_merger*big\_change)	
	}
	\Else{
		\tcp{if the merger vehicle is in front}
		\tcp{clamp desired merger velocity to make it coherent with driving in front of other}
		new\_vx\_merger = clamp(desired\_vx\_merger, min( old\_vx\_merger*big\_change,  old\_vx\_other/big\_change), old\_vx\_merger*big\_change)\\
		\tcp{also clamp the desired speed of the other, distinguishing between two cases:}
		\If{new\_vx\_merger $>$ old\_vx\_other}{
			new\_vx\_other = clamp(desired\_vx\_other, old\_vx\_other / small\_change, old\_vx\_other*small\_change)
		}
		\Else{
			new\_vx\_other = clamp(desired\_vx\_other, old\_vx\_other / big\_change, old\_vx\_other)
		}		
	}
\end{algorithm}

\emph{NB: on what follows, we refer to the method TGL-DP also simply by TGL-P:}

Let us give further details on our method \emph{TGL-P}, that was only briefly introduced in \Cref{sec:expdetails}.
This method is the same as TGL-D described in \Cpaperref{sec:pred} (building on \Cpaperref{sett:driving}), except that we modified the preference revelation net based on plausible reasoning and using parts of the equilibrium refinement/weighting net.
Note that \emph{this is still a special case of our general architecture (\Cpaperrefp{sec:archi}) in the rigorous sense},
just with a preference revelation net that incorporates a fair amount of additional reasoning and shares some structure with the equilibrium refinement net.

There are two reasons why we introduce TGL-P:
First,
the preference revelation net has to be trained together with the local game solver implicit layer.
This means that training takes comparably long (we have an outer and an inner optimization loop, as described in \Cpaperrefp{sec:archi}). And one particular problem we experienced and have not fully solved yet, is that often the preference revelation net already starts overfitting while the game parameters that are learned ``globally'', i.e., not inferred from the past trajectory by the preference revelation net, are not properly learned yet. 
Now TGL-P allows to demonstrate what we believe TGL itself can also achieve once the mentioned problems are overcome; 
in particular, it shows that the \emph{game model class} ($\game_\gparam$) has a substantial capacity to resemble the future trajecories.
Second, 
TGL-P shows how the intermediate representation $\gparam$ can be inspected and high level knowledge/reasoning about agents' preferences/utilities and behavior can be incoporated.

Specifically, TGL-P can be described as being the same as TGL, except that the part of preference revelation net of TGL that outputs $\gparam\ai{\nam{v}, 2}, \gparam\ai{\nam{v}, 1}$ is replaced by \Cref{alg:tglp} (on page \pageref{alg:tglp}). The function $clamp(\cdot)$ is, as usual, defined by
\begin{align*}
clamp(z,m,M) = \max ( m, \min(z,M)),
\end{align*}
i.e., clipping values to $m$ / $M$ if they are below / above.
While the details of the algorithm may look complex, the main idea is simple: we clip the outputs of the preference revelation net if they are too far off compared to what one would expect given initial velocities and output of the equilibrium refinement/weighting net.
The algorithm has two parameters $big\_change$ and $small\_change$, for which we used the values 1.2 and 1.04, respectively, in the experiment.

Note that, while in general, the equilibrium refinement/weighting net and the preference revelation net serve separate purposes, the reason why in TGL-P we share structure between them is mainly a pragmatic one: the equilibrium refinement/weighting net \emph{can} be trained separately from the implicit layer and thus much faster -- and we made the experience that it learns a reliable signal to predict the future (subspaces / selected equilibria). 

\subsection{Additional details and link for new highway data set HEE published alongside this paper}
\label{sec:expdata}

\begin{figure*}
	\centering
	\includegraphics[width=.6\linewidth]{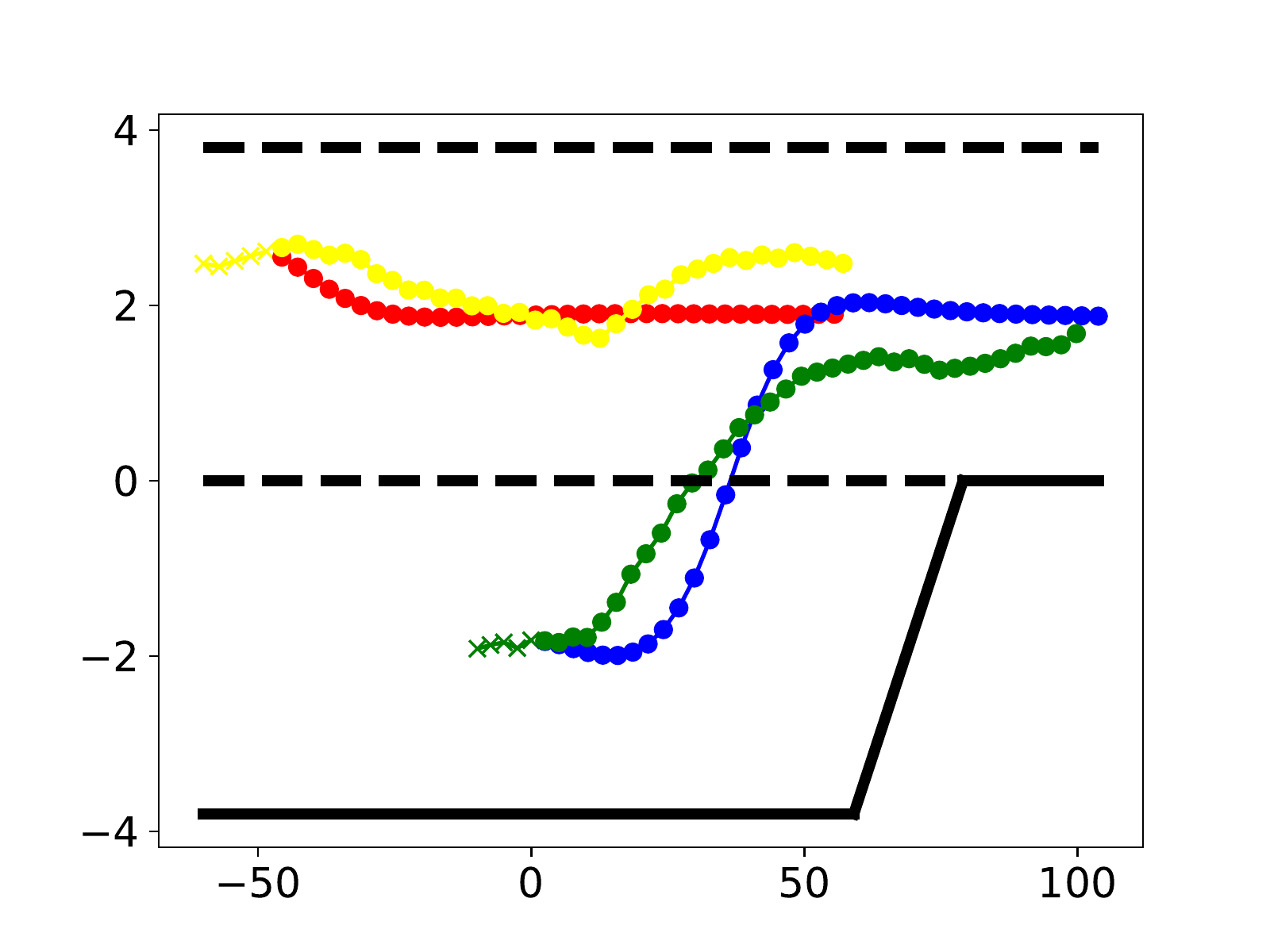}
	\includegraphics[width=.6\linewidth]{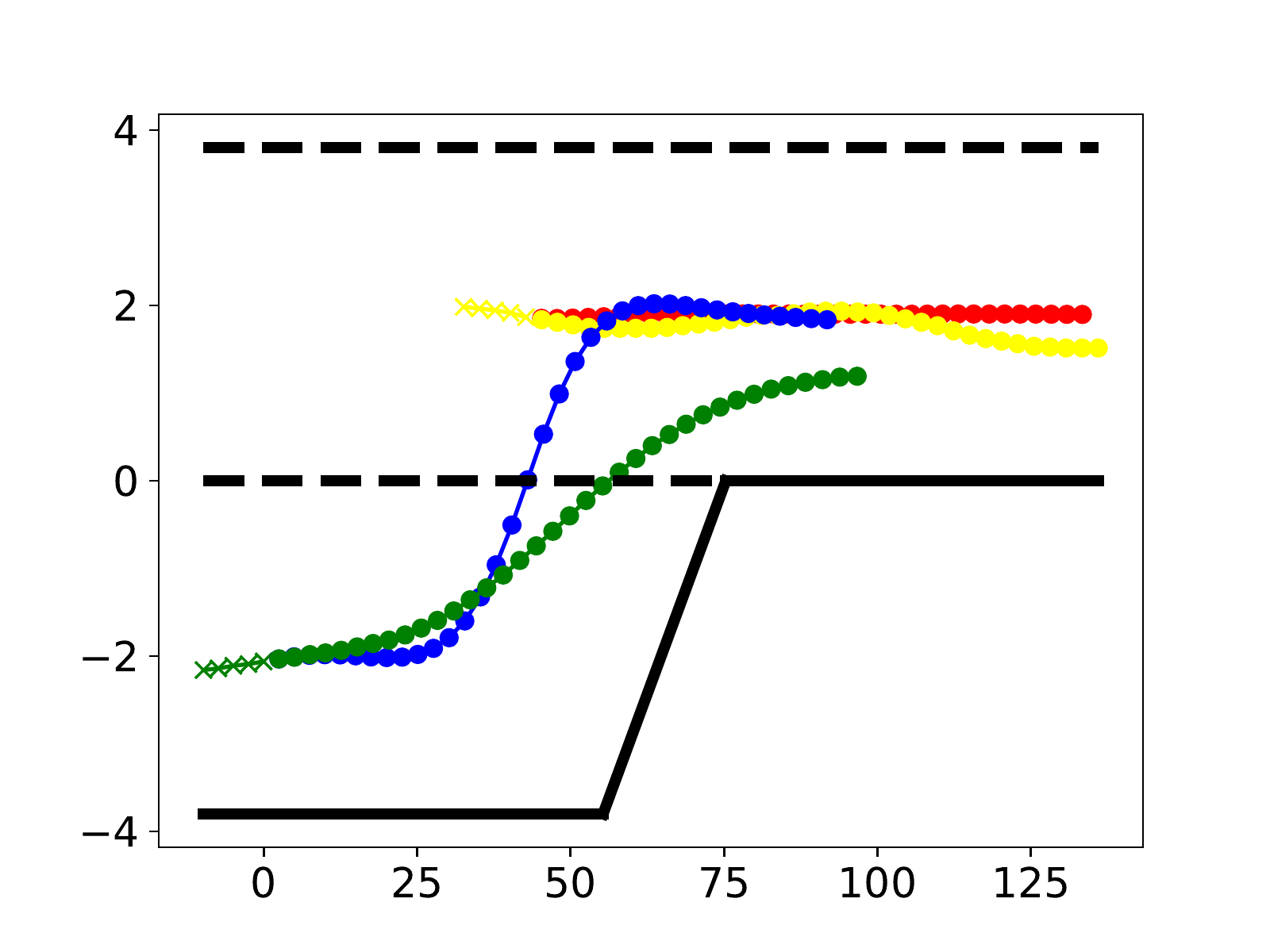}
	\caption{Example joint trajectories on \emph{highD data set (top)} and \emph{our new highway data set (bottom)}. The \emph{ground truth future} $\traj$ is green and yellow for on-ramp and highway car, respectively, with the past trajectory segment $\inittraj$ (at the very beginning) depicted by 'x' markers. The \emph{prediction $\ptraj$ (most likely mode) of our method TGL-P} is in blue and red, respectively. Note that x- and y-axis are in \emph{meters}; in particular, the \emph{x-axis is significantly squeezed}.}
	\label{fig:trajs}
\end{figure*}

One of the data sets we used in the experiment is a new one which we publish along this paper (introduced in \Cpaperrefp{sec:pred}).
We refer to it as \defi{HEE (Highway Eagle Eye) data set}.
\emph{Here are the links relevant for the data set:} 
\begin{itemize}
	\item Data set itself: \url{https://github.com/boschresearch/hee_dataset}.
	\item For example code of how to load/preprocess trajectories from the raw data set, please look at the code repository for this paper: \url{https://github.com/boschresearch/trajectory_games_learning}. Note that this repository also contains the precise \emph{filtered data set of two-player highway merge scenes that we used in the experiments}.
\end{itemize}
Further remarks on HEE:
\begin{itemize}
	\item The full recorded highway section in HEE is roughly as the lower lane (incl.\ exit/entry) in \Cref{fig:heefull} (on page \pageref{fig:heefull}) (the recorded section is in fact slightly more to the right than the picture indicates).
For the experiments we focused on merging scene trajectories, which roughly take place within the smaller highway section depicted in \Cref{fig:heemerge} (on page \pageref{fig:heemerge}).
Note that the images in \Cpaperrefp{fig:archi} are a stylized (significantly squashed in the x-dimension) version of \Cref{fig:heemerge}.
\item Note that the recorded highway section in our data set --  length $\sim600$m -- is longer compared to the one section with an on-ramp in the highD data set \citep{highDdataset}. But highD also contains other highway sections.
\item As stated earlier, keep in mind that this HEE data set can be useful for studies like ours, but some aspects of it may be noisy, \emph{so it is only meant for experimental purposes}.\footnote{One thing we found was that there is a slight mismatch between trajectory data compared to a preliminary map of the highway section (available in the data repository as well). This rather seems an error in the map, but could also be a distortion in the trajectory recordings.}


\end{itemize}

\subsection{More details on experimental results}
\label{sec:expres}

\begin{table}[!h]
	\renewcommand{\arraystretch}{1.2}  
	\begin{tabularx}{\linewidth}{Xcccccc} 
		\hlineB{2.5}
		Data set & Horizon  & TGL (ours) & TGL-D (ours) & TGL-DP (ours)  & CS-LSTM 
		& MFP 
		\\
		
		\hline 
		\multirow{2}{*}{highD \citep{highDdataset}}
& 1s & 0.5& 0.5& 0.5& 1.4& 1.2 \\ \cline{2-7} 
& 2s & 1.2& 1.1& 1.0& 2.4& 2.3 \\ \cline{2-7} 
& 3s & 2.1& 1.7& 1.6& 3.7& 3.6 \\ \cline{2-7} 
& 4s & 3.2& 2.6& 2.4& 4.8& 5.0 \\ \cline{2-7} 
& 5s & 4.5& 3.6& 3.2& 6.0& 6.6 \\ \cline{2-7} 
& 6s & 6.1& 4.8& 4.2& 7.0& 8.2 \\ \cline{2-7} 
& 7s & 7.9& 6.1& 5.4& 10.1& 9.8 \\ 
		
		\hline 
		\multirow{2}{*}{New data set (\Cref{sec:pred})}
& 1s & 0.8& 0.7& 0.8& 0.8& 0.8 \\ \cline{2-7} 
& 2s & 1.5& 1.4& 1.5& 1.6& 1.6 \\ \cline{2-7} 
& 3s & 2.4& 2.2& 2.3& 2.3& 2.4 \\ \cline{2-7} 
& 4s & 3.5& 3.1& 3.0& 3.3& 3.6 \\ \cline{2-7} 
& 5s & 4.6& 4.0& 3.9& 4.4& 4.6 \\ \cline{2-7} 
& 6s & 5.9& 4.9& 4.7& 5.3& 5.8 \\ \cline{2-7} 
& 7s & 7.3& 5.9& 5.6& 7.2& 7.1 \\ 
		
		\hlineB{2.5}
	\end{tabularx}
	\caption{Mean absolute error (MAE) for all prediction horizons for all methods.}
	\label{table:mae}
	%
	\renewcommand{\arraystretch}{1.2}  
	\begin{tabularx}{\linewidth}{Xcccccc} 
		\hlineB{2.5}
		Data set & Horizon  & TGL (ours) & TGL-D (ours) & TGL-DP (ours)  & CS-LSTM 
		& MFP 
		\\
		
		\hline 
		\multirow{2}{*}{highD \citep{highDdataset}}
& 1s & 0.6& 0.5& 0.5& 1.8& 1.6 \\ \cline{2-7} 
& 2s & 1.4& 1.2& 1.2& 3.2& 3.0 \\ \cline{2-7} 
& 3s & 2.7& 2.1& 2.0& 5.0& 4.8 \\ \cline{2-7} 
& 4s & 4.3& 3.2& 3.0& 6.3& 6.8 \\ \cline{2-7} 
& 5s & 6.1& 4.5& 4.2& 7.7& 9.0 \\ \cline{2-7} 
& 6s & 8.4& 6.1& 5.7& 9.1& 11.2 \\ \cline{2-7} 
& 7s & 11.0& 8.0& 7.6& 14.5& 13.6 \\
		
		\hline 
		\multirow{2}{*}{New data set (\Cref{sec:pred})}
 & 1s & 0.8& 0.7& 0.8& 0.9& 1.0 \\ \cline{2-7} 
& 2s & 1.8& 1.7& 1.7& 2.0& 1.9 \\ \cline{2-7} 
& 3s & 3.1& 2.8& 2.8& 2.8& 3.1 \\ \cline{2-7} 
& 4s & 4.4& 3.9& 3.9& 4.0& 4.5 \\ \cline{2-7} 
& 5s & 5.9& 5.2& 5.1& 5.1& 6.0 \\ \cline{2-7} 
& 6s & 7.6& 6.6& 6.4& 6.5& 7.6 \\ \cline{2-7} 
& 7s & 9.4& 8.1& 7.7& 8.7& 9.3 \\
		
		\hlineB{2.5}
	\end{tabularx}
	\caption{Root mean square error (RMSE) for all prediction horizons for all methods.}
	\label{table:rmse}
\end{table}

In the main text we only gave results averaged over all prediction horizons.
Here, Tables \ref{table:mae}, \ref{table:rmse} (on page \pageref{table:mae}) give the results -- MAE and RMSE for all methods (including CS-LSTM \citep{deo2018convolutional}, MFP \cite{tang2019multiple})  -- for \emph{each prediction horizon individually}, from 1s to 7s.

Additionally, in \Cref{fig:trajs} (on page \pageref{fig:trajs}) we show example joint trajectories on highD data set and our new highway data set: the past trajectory $\inittraj$, the prediction $\ptraj$ (most likely mode) of our method TGL-DP, and the ground truth future trajectory $\traj$.
Note that the \emph{x-axis is significantly squeezed} in these figures.

\section{Further related work}

\label{sec:morerelated}

Regarding work that combines machine learning and game theory, noteworthy is also \citep{hartford2016deep} who learn solution concepts, but not equilibrium refinement concepts. 
Generally, learning agent behavior, but in an active/experimental settings, is also extensively studied, e.g., in multiagent reinforcement learning \cite{shoham2008multiagent}.
Recently, also other related setting have been studied, e.g., where a computational assistant interacts with agents to learn their preferences and based on this support their coordination in the game-theoretic sense of equilibrium selection \citep{geiger2019coordination}.

Regarding imitation learning of agents' interaction from observational data when there may be relevant unobserved confounders, there is also work based on causal models with a focus on identifiability analysis \cite{etesami2020causal,causalconfusion,geiger2016experimental}.

Beyond inverse reinforcement learning mentioned already in the introduction, our work is closely related to the general area of imitation learning \citep{osa2018algorithmic}.

Regarding probabilistic (multi-)agent trajectory prediction, on the machine learning side there are also recent methods using normalizing flows \citep{bhattacharyya2019conditional,bhattacharyya2020haar}.

Regarding implicit layers, the following work is also worth mentioning explicitly:
\citet{amos2017optnet} take a (constrained) optimizer as implicit layer and derive the implicit layer's gradient formula from the (Lagrangian/Karush-Kuhn-Tucker) optimality condition. 
Note that this is similar to one part of our \Cref{thm:implicitlayer}, which, after the reduction to potential games, also derives the implicit layer's gradient formula from an optimality condition. But our \Cref{thm:implicitlayer} contains additional elements in the sense of the precise preconditions to fit with the rest of our setting, the parallelization of solving for several local optima/equilibria, and the continuity implication for the $\expli_k$, which follows easily but which, so far, we have not seen in other work.
Also note \citet{amos2018differentiable} who differentiate through (single-agent) model predictive control (MPC) conditions.
Worth mentioning are also \citet{li2020end}, who also use implicit differentiation to learn about games. 

\section{Remarks on rationality, local Nash equilibria, equilibrium selection, etc.}
\label{sec:ratio}

Let us make some high level remarks.

\begin{itemize}
	\item As a remark on the (local) rationality principle in our model: We do not think that humans are perfectly instrumentally rational (i.e., strategic utility maximizers) in general, and in many multiagent trajectory settings \emph{bounded} rationality and other phenomena may play an important role. Nonetheless, we believe that instrumental rationality is \emph{reasonably good approximation} in many settings. And of course, game-theoretic rationality has the advantage that there is an elegant theory for it.
	\item In fact we use the \emph{local} Nash equilibrium, which can also be seen as a bounded form of rationality. However, we feel that it can often be a better, more advanced approximation (to reality and/or rationality) than other concepts like level-k game theory \citep{stahl1995players}. In particular, we found it interesting that the local Nash equilibria can in fact correspond to intuitive \emph{modes} of the joint trajectories, like which car goes first in \Cpaperrefp{fig:archi}. A reason for this may be that, while there may be one perfect solution (say a global Nash), due to errors and stochasticity of the environment the agents may be perturbed towards ending up at a state where a previously local Nash equilibrium may in fact now be a global Nash equilibrium.
	\item The problem of equilibrium selection mentioned in the main text may be seen as a form of \emph{incompleteness} of game theory -- it cannot always make a unique prediction (or prescription). Our equilibrium refinement and weighting nets can be seen as a data-driven approach to fill this incompleteness. An interesting question in this context is whether the missing information is actually contained in the preferences of the agents, or if there is additional (hidden) information required to find the unique solution/prediction.
	\item 
	In some situations, it may be that knowledge of an (local-)equilibrium-separating partition may in fact tractably tell us the \emph{global} Nash equilibrium (because if we know all local optima then we can also know the global optimum of a function). But this does not always seem to be the case for general common-coupled/potential games and, in particular, also because we do not require the equilibrium-separation to cover all possible equilibria.
\end{itemize}